\documentclass[a4paper,onecolumn,accepted=2026-01-04]{quantumarticle}
\pdfoutput=1

\usepackage[utf8]{inputenc}
\usepackage[margin=1in]{geometry}
\usepackage{color, colortbl}
\usepackage[dvipsnames]{xcolor}
\usepackage{graphicx,epstopdf}
\usepackage{amsmath,amssymb,amsthm}
\usepackage{algorithm}
\usepackage{algorithmic}
\usepackage{bm}
\usepackage[caption=false]{subfig}
\usepackage{appendix}
\usepackage{multirow}
\usepackage{mathtools}
\usepackage{braket}
\usepackage{csquotes}
\usepackage[sort,square,numbers]{natbib}
\usepackage[colorlinks=true,linkcolor=red,bookmarks=true,breaklinks=true]{hyperref}
\usepackage[english]{babel}
\usepackage[qm]{qcircuit}
\usepackage[capitalize,nameinlink]{cleveref}
\usepackage{tikz}
\usetikzlibrary{shapes.geometric}
\usetikzlibrary{arrows.meta}
\usetikzlibrary{positioning}
\usetikzlibrary{calc}
\usepackage{array}
\usepackage[normalem]{ulem}
\usepackage{makecell}
\usepackage{authblk}

\definecolor{color1}{HTML}{FFDAB9}
\definecolor{color2}{HTML}{FFA07A}
\definecolor{color3}{HTML}{FF7F50}
\definecolor{color4}{HTML}{FF6347}

\definecolor{amber}{rgb}{1.0, 0.75, 0.0}

\renewcommand{\d}{\mathrm{d}}

\newtheorem{thm}{\protect\theoremname}
  \theoremstyle{plain}
  \newtheorem{lem}[thm]{\protect\lemmaname}
  \theoremstyle{remark}
  
  \theoremstyle{plain}
  \newtheorem*{lem*}{\protect\lemmaname}
  \theoremstyle{plain}
  
  \theoremstyle{plain}
  \newtheorem{cor}[thm]{\protect\corollaryname}

  \newtheorem{result}[thm]{Result}
  
  \newcolumntype{x}[1]{>{\centering\let\newline\\\arraybackslash\hspace{0pt}}p{#1}}
  
\PassOptionsToPackage{USenglish}{babel}
\usepackage[USenglish]{babel}
  \providecommand{\corollaryname}{Corollary}
  \providecommand{\lemmaname}{Lemma}
  \providecommand{\propositionname}{Proposition}
  \providecommand{\remarkname}{Remark}
\providecommand{\theoremname}{Theorem}

\newcommand{\norm}[1]{\left\lVert#1\right\rVert}

\newcommand{\Or}{\mathcal{O}}

\let\originalleft\left
\let\originalright\right
\renewcommand{\left}{\mathopen{}\mathclose\bgroup\originalleft}
\renewcommand{\right}{\aftergroup\egroup\originalright}

\begin{document}

\title{Fast-forwarding quantum algorithms for linear dissipative differential equations}

\author{Dong An}
\affiliation{Beijing International Center for Mathematical Research (BICMR), Peking University, Beijing, China}
\affiliation{Joint Center for Quantum Information and Computer Science (QuICS), University of Maryland, College Park, MD, USA}
\author{Akwum Onwunta}
\affiliation{Department of Industrial and Systems Engineering, Lehigh University, Bethlehem, PA, USA}
\author{Gengzhi Yang}
\affiliation{Joint Center for Quantum Information and Computer Science (QuICS), University of Maryland, College Park, MD, USA}
\affiliation{Department of Mathematics, University of Maryland, College Park, MD, USA}


\maketitle

\begin{abstract}
    We establish improved complexity estimates of quantum algorithms for linear dissipative ordinary differential equations (ODEs) and show that the time dependence can be fast-forwarded to be sub-linear. Specifically, we show that a quantum algorithm based on truncated Dyson series can prepare history states of dissipative ODEs up to time $T$ with cost $\widetilde{\mathcal{O}}(\log(T) (\log(1/\epsilon))^2 )$, which is an exponential speedup over the best previous result. For final state preparation at time $T$, we show that its complexity is $\widetilde{\mathcal{O}}(\sqrt{T} (\log(1/\epsilon))^2 )$, achieving a polynomial speedup in $T$. We also analyze the complexity of simpler lower-order quantum algorithms, such as the forward Euler method and the trapezoidal rule, and find that even lower-order methods can still achieve $\widetilde{\mathcal{O}}(\sqrt{T})$ cost with respect to time $T$ for preparing final states of dissipative ODEs. As applications, we show that quantum algorithms can simulate dissipative non-Hermitian quantum dynamics and heat processes with fast-forwarded complexity sub-linear in time. 
\end{abstract}

\tableofcontents

\section{Introduction}

Differential equations are fundamental in modeling the dynamics of a system and have been widely used in physics, chemistry, engineering, economics, and other disciplines. In this paper, we 
consider the solution of a system of ordinary differential equations (ODEs) of the form
\begin{align}
    \frac{\d u(t)}{\d t} &= A(t) u(t) + b(t), \quad t \in [0,T], \label{eqn:ODE}\\
    u(0) &= u_0. 
\end{align}
Here, $A(t) \in \mathbb{C}^{N\times N}$ and $b(t) \in \mathbb{C}^{N}$. 
We are interested in approximating the final solution $u(t)$ at time $t=T$, 
or computing a history state encoding the trajectory of the dynamics, i.e., $\mathbf{u} = [u_0;u_1;\cdots;u_M]^\top$ such that $u_j \approx u(jT/M)$ where $M$ is the number of time steps, and we let $h = T/M$ denote the time step size. 

In practice, the dimension $N$ of~\cref{eqn:ODE} can be huge, especially in semi-discretized partial differential equations with high accuracy or systems that model the motion and interaction of a large number of objects. 
This high dimensionality poses significant computational challenges for efficient classical simulation of~\cref{eqn:ODE}, and has evoked emerging interest in exploring quantum algorithms for differential equations with substantially improved efficiency. 

The first efficient quantum algorithm for general ODEs can be dated back to~\cite{Berry2014}, which leverages the power of quantum linear system algorithms. 
Specifically, the algorithm in~\cite{Berry2014} first discretizes~\cref{eqn:ODE} by a multi-step method, formulates the discretized ODE as a linear system of equations, and solves the linear system by the Harrow–Hassidim–Lloyd (HHL) algorithm~\cite{HarrowHassidimLloyd2009}.  
Since then, there have been several subsequent works on improved quantum ODE algorithms based on the same idea of converting the ODE to a linear system of equations but with more advanced time discretization methods and quantum linear system algorithms~\cite{BerryChildsOstranderEtAl2017,ChildsLiu2020,Krovi2022,BerryCosta2022}. 
There have also been attempts on designing quantum ODE algorithms based on ideas other than the linear-system-based approach, such as time-marching~\cite{FangLinTong2022}, quantum eigenvalue processing~\cite{lowSu2024}, and reduction to Hamiltonian simulation problems~\cite{AnLiuLin2023,AnChildsLin2023,JinLiuYu2022,HuJinLiuEtAl2023}.

Since simulating ODEs for a long time period appears ubiquitously in scientific and engineering computation, an important aspect among others of quantum ODE algorithms is how the complexity depends on the evolution time $T$. 
The state-of-the-art linear-system-based algorithm is the truncated Dyson series method~\cite{BerryCosta2022}, which uses the truncated Dyson series for time discretization, applies optimal quantum linear system algorithm~\cite{CostaAnYuvalEtAl2022}, and achieves $\widetilde{\Or}( T )$ query complexity. 
Similarly, the best available method without solving a linear system of equations also has query complexity $\widetilde{\Or}( T )$~\cite{AnChildsLin2023}. 
Both of the two strategies scale almost linearly in the evolution time $T$, which can still be expensive for long time simulation.

In this work, we explore the possibility of further fast-forwarding quantum ODE algorithms in time dependence and focus on the following question: 
\begin{quote} 
\textit{Can we solve (a subset of) differential equations on a quantum computer with a sub-linear cost in the evolution time?} 
\end{quote}
Notice that for Hamiltonian simulation problems where $A(t)$ is anti-Hermitian and $b(t) \equiv 0$,~\cite{BerryChildsCleveEtAl2014} establishes a lower bound with linear time dependence $\Omega(T)$, so we do not expect such fast-forwarding to be generically possible, though fast-forwarded Hamiltonian simulation of specific systems has been proposed in~\cite{AtiaAharonov2017,GuSommaSahinoglu2021}. 
Instead, in this work we focus on dissipative ODEs where $A(t)$ has a uniformly negative logarithmic norm. 

For dissipative ODEs, we establish an improved complexity estimate for all linear-system-based algorithms regardless of time discretization scheme. 
When applied to the truncated Dyson series method, our analysis yields an $\widetilde{\Or}(\text{poly}\log(T/\epsilon))$ query complexity for history state preparation. 
This is an exponential improvement in evolution time $T$ compared to existing results. 
For final state preparation, we show that the truncated Dyson series method can scale only $\widetilde{\Or}(\sqrt{T}\text{~poly}\log(1/\epsilon))$, achieving a quadratic speedup over the best existing results. 
We also analyze the complexity of linear-system-based algorithms with simpler lower-order time discretization schemes, including the forward Euler method and trapezoidal rule. 
Interestingly, we find that with lower-order methods we can still achieve $\widetilde{\Or}(\sqrt{T})$ for final state preparation in terms of $T$, and lower convergence order only worsens the error dependence. 
Our algorithms and analysis can be applied to simulating dissipative quantum dynamics with non-Hermitian Hamiltonian and heat process.

\subsection{Main results}

We study fast-forwarding quantum algorithms for linear dissipative ODEs. 
In~\cref{eqn:ODE}, we assume the logarithmic norm of $A(t)$ is uniformly negative, i.e., there exists a constant $\eta > 0$ such that 
\begin{equation}
    A(t) + A(t)^{\dagger} \leq -2\eta < 0. 
\end{equation}
We are interested in computing a final state proportional to $u(T)$, or a history state proportional to 
\begin{equation}
    \mathbf{u}_{\mathrm{exact}} = [u(0);u(T/M);u(2T/M);\cdots;u(T)],
\end{equation}
where $M$ is the number of time steps. 
Our main mathematical result is a general quantum complexity estimate for dissipative ODEs. 

\begin{result}[Informal version of~\cref{thm:complexity_single_step_inhomo,cor:complexity_final_state_optimal_Mp}]\label{res:complexity_general}
    Consider~\cref{eqn:ODE} with $A(t)+A(t)^{\dagger} \leq -2\eta < 0$ for a constant $\eta > 0$ and we use $M$ steps for time discretization. 
    Then, we can prepare an $\epsilon$-approximation of 
    \begin{enumerate}
        \item the history state $\mathbf{u}_{\mathrm{exact}}/\norm{\mathbf{u}_{\mathrm{exact}}}$ up to time $T$ with query complexity 
        \begin{equation}
            \Or\left(\frac{M}{T}\log\left(\frac{1}{\epsilon}\right)\right), 
        \end{equation}
        \item  the final state $u(T)/\norm{u(T)}$ at time $T$ with query complexity 
        \begin{equation}
            \widetilde{\Or}\left( \frac{\max_t\norm{u(t)}}{\norm{u(T)}} \frac{M}{\sqrt{T}} \log\left( \frac{1}{\epsilon} \right) \right). 
        \end{equation}
    \end{enumerate}
\end{result}

The algorithm used in~\cref{res:complexity_general} is a slightly modified linear-system-based approach from~\cite{Berry2014,BerryChildsOstranderEtAl2017,ChildsLiu2020,Krovi2022,BerryCosta2022}. 
Specifically, for history state preparation, we use $M$ many equidistant time steps to discretize~\cref{eqn:ODE}. 
Since~\cref{eqn:ODE} is linear, its discretized version is naturally a linear system of equations with solution $\mathbf{u} = [u_0;u_1;u_2;\cdots;u_M]$ where each $u_j$ aims to approximate $u(jT/M)$. 
Then we solve this linear system of equations by quantum algorithms with optimal asymptotic scaling (e.g.,~\cite{LinTong2020,CostaAnYuvalEtAl2022,Dalzell2024}) to prepare the desired history state. 
Notice that a final state $u(T)/\norm{u(T)}$ can be directly obtained from the history state by post-selecting the register encoding the index of the time step onto $M$. 
However, the success probability can be small as most blocks in the history state are intermediate steps. 
To boost the success probability, we may use the padding trick introduced in~\cite{Berry2014} by adding trivial lines at the end of the linear system of equations to make multiple copies of the final state. 
A difference from~\cite{Berry2014} is that we can use much fewer padding lines for dissipative ODEs, and we will discuss the reason later. 
A detailed discussion on the quantum algorithms is presented in~\cref{sec:algorithm}.

For history state preparation, our~\cref{res:complexity_general} shows that the complexity is only $\Or\left( \frac{M}{T} \log\left(\frac{1}{\epsilon}\right) \right)$, where the factor $\Or(M/T)$ is the condition number of the linear system, and the $\Or(\log(1/\epsilon))$ is due to the optimal quantum linear system algorithm. 
Compared to the previous analysis, our improvement lies in a better estimate of the condition number. 
Specifically, previous analysis~\cite{Berry2014,BerryChildsOstranderEtAl2017,ChildsLiu2020,Krovi2022,BerryCosta2022} only bounds the condition number by $\Or(M)$, so we gain a multiplicative factor of $\Or(1/T)$. 
Such an improvement is crucially dependent on the additional dissipation in the dynamics. 
Recall that the condition number of the linear system represents how a small perturbation in the right-hand side is amplified in the solution. 
For a stable but non-dissipative dynamics\footnote{This refers to the dynamics~\cref{eqn:ODE} where $A(t)$ has only non-positive logarithmic norm, i.e., $A(t)+A(t)^{\dagger} \leq 0$, instead of the dissipative ODE which requires the logarithmic norm to be strictly smaller than $0$. Notice that the stable condition $A(t)+A(t)^{\dagger} \leq 0$ is also assumed in previous works for quantum ODE algorithms~\cite{BerryCosta2022,AnLiuLin2023,AnChildsLin2023} and includes the case of Hamiltonian simulation.}, a small perturbation in the initial condition $u_0$ or the inhomogeneous term $b(t)$ can be carried out to the entire dynamics with the same amplitude, so the condition number becomes linear in terms of the number of steps. 
However, in a dissipative ODE, the effect of a small perturbation in $u_0$ or $b(t)$ decays exponentially in time, so such perturbation will only introduce non-trivial perturbations on the dynamics for a fixed time period instead of the entire time period up to $T$. 
This is why in the condition number we can obtain a $\Or(1/T)$ multiplicative factor improvement.

For final state preparation, our~\cref{res:complexity_general} gives an $\widetilde{\Or}\left( \frac{\max_t\norm{u(t)}}{\norm{u(T)}} \frac{M}{\sqrt{T}} \log\left( \frac{1}{\epsilon} \right) \right)$ query complexity, while previous analysis in~\cite{Berry2014,BerryChildsOstranderEtAl2017,ChildsLiu2020,Krovi2022,BerryCosta2022} suggests the query complexity to be $\widetilde{\Or}\left( \frac{\max_t\norm{u(t)}}{\norm{u(T)}} M \log\left( \frac{1}{\epsilon} \right) \right)$. 
Unlike the history state preparation case, here we only obtain an $\Or(1/\sqrt{T})$ factor of improvement. 
This is because of the fewer padding lines we use in the linear system. 
Specifically, notice that if we add $M_p$ many padding lines, then the condition number will be increased by an additive factor $\Or(M_p)$. 
In the general stable but non-dissipative case, the condition number of the linear system without padding is $\Or(M)$, so we can add as many as $M_p = M$ padding lines while not increasing the condition number asymptotically. 
Then the number of copies of the final state $u(T)$ is the same as the number of intermediate steps $u(jT/M)$ for $j < M$, so the probability of extracting $u(T)$ from the padded history state only depends on the norm decay rate $\max_t \norm{u(t)}/\norm{u(T)}$ and the overall complexity is still linear in $\Or(M)$. 
However, for the dissipative ODEs, we have shown that the condition number without padding is just $\Or(M/T)$, so the maximum number of extra padding lines allowed without increasing the condition number asymptotically is just $M_p \sim M/T$, which is asymptotically smaller than the number of time steps. 
Then we need more repeats to successfully extract the final state $u(T)$ from the padded history state, and with amplitude amplification, this cancels an $\Or(\sqrt{T})$ factor. 
We remark that the choice $M_p \sim M/T$ is indeed optimal: we show in~\cref{thm:complexity_final_state} the query complexity of final state preparation for any choice of $M_p$ and find the optimal choice to be $\Or(M/T)$ in~\cref{cor:complexity_final_state_optimal_Mp}.

So far we have not specified a concrete time discretization method in the algorithm. 
In fact, our~\cref{res:complexity_general} works for any single step time discretization method, including the simplest forward Euler method, Runge-Kutta method, and the truncated Dyson series method, while the choice of $M$ to achieve a certain level of accuracy varies among different methods. 
For the truncated Dyson series method~\cite{BerryCosta2022}, our analysis yields the following complexity for dissipative ODEs. 

\begin{result}[Informal version of~\cref{cor:Dyson_hist_inhomo}]\label{res:Dyson}
    Consider the truncated Dyson series method for~\cref{eqn:ODE} with $A(t)+A(t)^{\dagger} \leq -2\eta < 0$. 
    Then, we can prepare an $\epsilon$-approximation of 
    \begin{enumerate}
        \item the history state $\mathbf{u}_{\mathrm{exact}}/\norm{\mathbf{u}_{\mathrm{exact}}}$ up to time $T$ with query complexity
        \begin{equation}
            \Or\left(\log(T)\left(\log\left(\frac{1}{\epsilon}\right)\right)^2\right), 
        \end{equation}
        \item  the final state $u(T)/\norm{u(T)}$ at time $T$ with query complexity
        \begin{equation}
            \widetilde{\Or}\left( \frac{\max_t\norm{u(t)}}{\norm{u(T)}} \sqrt{T} \left(\log\left(\frac{1}{\epsilon}\right)\right)^2 \right). 
        \end{equation}
    \end{enumerate}
\end{result}

The proof of~\cref{res:Dyson} is discussed in~\cref{sec:Dyson}, which we can obtain from our general~\cref{res:complexity_general} by noticing that $M = \Or(T)$ is sufficient in the truncated Dyson series method. 
The extra $\log(T)$ factor in the history state preparation cost is due to the truncated order of the Dyson series, which typically depends on the evolution time and requires more queries to the matrix $A(t)$ for constructing the linear system. 
The highlight of~\cref{res:Dyson} is the time dependence: in both history and final state preparation cases, we obtain sub-linear scalings in time $T$ and this is better than previous results. 
We will make a careful comparison between~\cref{res:Dyson} and existing works later in~\cref{sec:intro_comparison}.

\subsection{Additional results}

\paragraph{Lower-order methods: } In~\cref{res:Dyson}, we have analyzed the query complexity of the truncated Dyson series method for dissipative ODEs. 
Although the truncated Dyson series method can achieve high-order convergence, its quantum implementation requires complicated quantum control logic. 
In~\cref{sec:lower_order}, we also discuss the complexity of two lower-order methods that can be implemented in a much simpler way: the first-order forward Euler method and the second-order trapezoidal rule. 

\begin{result}\label{res:lower_order}
     Consider the forward Euler method and the trapezoidal rule for~\cref{eqn:ODE} with $A(t)+A(t)^{\dagger} \leq -2\eta < 0$. 
     Then, we can prepare an $\epsilon$-approximation of 
     \begin{enumerate}
        \item the history state $\mathbf{u}_{\mathrm{exact}}/\norm{\mathbf{u}_{\mathrm{exact}}}$ up to time $T$ with query complexity\footnote{For technical simplicity, here we only focus on the explicit dependence of $T$ and $\epsilon$, and assume all the other parameters to be constants (e.g., the decay rate $\max_t\norm{u(t)}/\norm{u(T)}$, the norms $\max_t\norm{A(t)}$ and $\norm{b(t)}$, and parameters related to the stability condition). The same also applies for the final state preparation. }
        \begin{equation}
            \Or\left(  \frac{T^{1/(2p)}}{\epsilon^{1/p}} \log\left(\frac{1}{\epsilon}\right)\right), 
        \end{equation}
        where $p = 1$ for the forward Euler, and $p=2$ for the trapezoidal rule, 
        \item  the final state $u(T)/\norm{u(T)}$ at time $T$ with query complexity
        \begin{equation}
            \widetilde{\Or}\left(\frac{\sqrt{T}}{\epsilon^{1/p}} \log\left(\frac{1}{\epsilon}\right) \right), 
        \end{equation}
        where $p = 1$ for the forward Euler, and $p=2$ for the trapezoidal rule. 
     \end{enumerate}
\end{result}

From~\cref{res:lower_order}, we can see that the error dependence is standard, i.e., $\Or(\epsilon^{-1/p})$ for a $p$-th order method, but the time dependence is always sub-linear. 
For history state preparation, higher-order method can provide a further improved time scaling. 
Interestingly, for final state preparation, both methods scale $\Or(\sqrt{T})$, which is exactly the same as the truncated Dyson series method. 
Therefore, if we only care about complexity in terms of time $T$, the simplest Euler method suffices to achieve the best available scaling. 
The reason why we can achieve the same time scaling is, although the condition number of the linear system becomes larger for lower-order methods as suggested by the complexity of history state preparation, we may correspondingly add more padding lines to increase the probability of measuring onto the final state without further increasing the condition number, so the final state preparation complexity can still remain asymptotically the same. 
We also expect~\cref{res:lower_order} to hold for general $p$-th order time discretization method.

\paragraph{Homogeneous ODEs: } So far, all the presented results hold for possibly inhomogeneous ODEs. 
Here we consider the special homogeneous case where $b(t) \equiv 0$. 
Notice that the final state preparation problem for homogeneous dissipative ODEs always has exponential dependence on the evolution time $T$, since the solution $u(t)$ decays exponentially and extracting an exponentially small component from a quantum state requires exponential cost~\cite{AnLiuWangEtAl2022}. 
So we only consider the history state preparation and have the following result. 

\begin{result}\label{res:homogeneous}
    Consider~\cref{eqn:ODE} with $A(t)+A(t)^{\dagger} \leq -2\eta < 0$ and $b(t) \equiv 0$. 
    Then, for preparing the history state $\mathbf{u}_{\mathrm{exact}}/\norm{\mathbf{u}_{\mathrm{exact}}}$ up to time $T$ with error at most $\epsilon$, 
    \begin{enumerate}
        \item the truncated Dyson series method has query complexity $\Or\left( \left( \log\left( \frac{1}{\epsilon} \right) \right)^2 \right)$, 
        \item the forward Euler method has query complexity $\Or\left( \frac{1}{\epsilon}\log\left( \frac{1}{\epsilon} \right)  \right)$, and  the trapezoidal rule has query complexity $\Or\left( \frac{1}{\sqrt{\epsilon}}\log\left( \frac{1}{\epsilon} \right)  \right)$. 
    \end{enumerate}
\end{result}

\cref{res:homogeneous} suggests that we can completely get rid of any explicit time dependence for homogeneous dissipative ODEs, no matter which time discretization method we use. 
Intuitively, this is because the solution of a homogeneous dissipative ODE decays exponentially in time, so the non-trivial part in a history state is always within a constant time period.

\subsection{Related works and comparison}\label{sec:intro_comparison}

\begin{table}[t]
    \renewcommand{\arraystretch}{1.8}
    \centering
    \scalebox{0.81}{
    \begin{tabular}{c|c|c|c|c}\hline\hline
        \multirow{2}{4em}{\textbf{Methods}} & \multicolumn{2}{c|}{\textbf{Assumptions}} & \multicolumn{2}{c}{\textbf{Query complexities}} \\
        \cline{2-5} & Time & Stability & History state & Final state \\\hline
        $p$-th order multi-step~\cite{Berry2014} & TI & $\mathrm{Re}(\mathrm{eig}(A)) \leq 0$ 
        & $\widetilde{\Or}\left( T^{2+2/p} / \epsilon^{1+2/p} \right)$  & $\widetilde{\Or}\left( T^{2+2/p} / \epsilon^{2} \right)$ \\
        Taylor~\cite{BerryChildsOstranderEtAl2017} & TI & $\mathrm{Re}(\mathrm{eig}(A)) \leq 0$ 
        & $\widetilde{\Or}\left( T \text{~poly}\log(1/\epsilon) \right)$  & $\widetilde{\Or}\left( T \text{~poly}\log(1/\epsilon) \right)$ \\
        Improved Taylor~\cite{Krovi2022} & TI & $\max_t \norm{e^{At}} = \Or(1)$ 
        & $\widetilde{\Or}\left( T \text{~poly}\log(1/\epsilon) \right)$  & $\widetilde{\Or}\left( T \text{~poly}\log(1/\epsilon) \right)$ \\\hline
        Spectral~\cite{ChildsLiu2020} & TD & $\mathrm{Re}(\mathrm{eig}(A(t))) \leq 0$ 
        & $\widetilde{\Or}\left( T \text{~poly}\log(1/\epsilon) \right)$  & $\widetilde{\Or}\left( T \text{~poly}\log(1/\epsilon) \right)$ \\
        Dyson~\cite{BerryCosta2022} & TD & $A(t)+A(t)^{\dagger} \leq 0$ 
        & $\widetilde{\Or}\left( T (\log(1/\epsilon))^2 \right)$  & $\widetilde{\Or}\left( T (\log(1/\epsilon))^2 \right)$ \\\hline
        Eigenvalue processing~\cite{lowSu2024} & TI & $ A+A^{\dagger} \leq 0$ 
        & / & $\Or((T+\log(1/\epsilon))\log(1/\epsilon))$ \\
        Time-marching~\cite{FangLinTong2022} & TD & $A(t)+A(t)^{\dagger} \leq 0$ 
        & / & $\widetilde{\Or}\left( T^2 \log(1/\epsilon) \right)$ \\
        Schr\"odingerization~\cite{JinLiuYu2022} & TD & $A(t)+A(t)^{\dagger} \leq 0$ 
        & / & $\widetilde{\Or}\left( T/\epsilon \right)$ \\
        LCHS~\cite{AnChildsLin2023} & TD & $A(t)+A(t)^{\dagger} \leq 0$ 
        & / & $\widetilde{\Or}\left( T (\log(1/\epsilon))^{2+o(1)} \right)$ \\\hline
        Fast-forwarded Taylor~\cite{JenningsLostaglioLowrieEtAl2024} & TI & $PA+A^{\dagger}P < 0 $ 
        & $\widetilde{\Or}\left( T^{1/2} (\log(1/\epsilon))^2 \right)$  & $\widetilde{\Or}\left( T^{3/4} (\log(1/\epsilon))^2 \right)$  \\
        Fast-forwarded Euler (\textcolor{red}{this work}) & TD & $A(t)+A(t)^{\dagger} < 0$ 
        & $\Or\left( T^{1/2} / \epsilon \right)$  & $\Or\left( T^{1/2} / \epsilon \right)$  \\
        Fast-forwarded Dyson (\textcolor{red}{this work}) & TD & $A(t)+A(t)^{\dagger} < 0$ 
        & $\Or\left( \log(T) (\log(1/\epsilon))^2 \right)$ & $\widetilde{\Or}\left( T^{1/2} (\log(1/\epsilon))^2 \right)$\\\hline\hline
    \end{tabular}
    }
    \caption{ A comparison among our fast-forwarded results and previous works. Here in the ``Assumptions: Time'' column, ``TI'' means ``time-independent''(the algorithm only applies to ODEs with time-independent matrix $A(t) \equiv A$), and ``TD'' means ``time-dependent'' (the algorithm works for ODEs with possibly time-dependent matrix $A(t)$). The ``Assumptions: Stability'' column indicates further stability condition on $A(t)$ for the algorithm to be effective. In the query complexities, we only present explicit scalings in time $T$ and error $\epsilon$, and assume all the other parameters to be constants (e.g., the decay rate $\max_t\norm{u(t)}/\norm{u(T)}$, the norm $\max_t\norm{A(t)}$, and parameters related to the stability condition). 
    }
    \label{tab:comparison}
\end{table}

As we discussed earlier, several generic quantum ODE algorithms have been proposed during the past decade. 
They can be roughly divided into two categories. 
The first category is the linear-system-based approach~\cite{Berry2014,BerryChildsOstranderEtAl2017,ChildsLiu2020,Krovi2022,BerryCosta2022}, which discretizes the dynamics in time, formulates the discretized ODE as a linear system of equations, and solves the linear system by quantum linear system solvers. 
The second category includes those that attempt to directly implement the time evolution operator of~\cref{eqn:ODE}~\cite{FangLinTong2022,JinLiuYu2022,AnLiuLin2023,AnChildsLin2023,lowSu2024}, and we call them evolution-based approach. 
Our algorithm in this work is almost the same as the linear-system-based approach only with a slight difference in the number of padding lines for final state preparation, and our contribution mainly lies in the improved complexity analysis. 

In~\cref{tab:comparison}, we compare query complexities of our fast-forwarded Dyson and Euler methods with these generic quantum ODE algorithms. 
The first group in~\cref{tab:comparison} is the linear-system-based approach for ODEs with time-independent coefficient matrix $A(t) \equiv A$, the second group is the linear-system-based approach for time-dependent $A(t)$, and the third group is the evolution-based approach. 
In previous works, it was not clearly stated how the evolution-based approach can efficiently prepare a history state, so we exclude them from the comparison of history state preparation. 
Notice that all the previous generic quantum ODE algorithms require additional stability conditions, specified in the third column of~\cref{tab:comparison}. 
Though technically different, these ensure the dynamics to be stable but not necessarily strictly dissipative, i.e., the homogeneous time evolution operator $\mathcal{T}e^{\int_0^t A(s) ds}$ does not increase asymptotically for long time. 
Our results require stronger stability condition that the dynamics needs to be strictly dissipative, but we can obtain improved time dependence. 
Compared with the best previous method, for history state preparation, our fast-forwarded results achieve an exponential improvement ($\Or(\log(T))$ versus $\widetilde{\Or}(T)$), and for final state preparation, our fast-forwarded results achieve a quadratic speedup ($\widetilde{\Or}(\sqrt{T})$ versus $\widetilde{\Or}(T)$).

Other than generic algorithms, there have been a few attempts at fast-forwarding algorithms for special types of ODEs. 
An earlier work~\cite{JenningsLostaglioLowrieEtAl2024} also fast-forwards linear-system-based quantum ODE algorithms for dissipative ODEs by analyzing the condition number of the linear system. 
We compare our results with~\cite{JenningsLostaglioLowrieEtAl2024} in the last group of~\cref{tab:comparison}, and there are two main differences. 
First, the applicable regimes are different.
\cite{JenningsLostaglioLowrieEtAl2024} considers the time-independent matrix $A$ and imposes the Lyapunov stability condition that there exists an invertible matrix $P$ such that $PA + A^{\dagger}P < 0$. 
Our results generalize the fast-forwarding results to the time-dependent matrix $A(t)$ but require a more restrictive stability on the negative logarithmic norm. 
Second, within the commonly applicable regime, our results give a tighter upper bound of the linear system condition number and yield better complexity estimates in terms of $T$. 
Specifically, for history state preparation, we can achieve $\Or(\log(T))$ complexity while~\cite{JenningsLostaglioLowrieEtAl2024} only has $\Or(\sqrt{T})$. 
For final state preparation, our results also improve the complexity from $\widetilde{\Or}(T^{3/4})$ to $\widetilde{\Or}(T^{1/2})$. 
Besides~\cite{JenningsLostaglioLowrieEtAl2024}, some other previous works also consider fast-forwarding quantum ODE algorithms but under different scenarios, such as~\cite{AtiaAharonov2017,GuSommaSahinoglu2021} for Hamiltonian simulation and~\cite{AnLiuWangEtAl2022} for imaginary time evolution. 

\subsection{Applications}

We have discussed quantum algorithms with fast-forwarded complexity for linear dissipative ODEs. 
Such linear dissipative ODEs have many applications including quantum dynamics with non-Hermitian Hamiltonian, heat processes, damped systems, and linearized dissipative nonlinear differential equations. 

In~\cref{sec:applications}, we discuss two applications in more details. 
The first application is the dissipative quantum dynamics with non-Hermitian Hamiltonian, which is described by a time-dependent Schr\"odinger equation with a non-Hermitian matrix as the Hamiltonian and has appeared ubiquitously in various branches of physics and chemistry, such as open quantum system, quantum resonances, quantum transport, field theories, quantum many-body system, to name a few~\cite{Bender2007,RegoMonteiroNobre2013,GiusteriMattiottiCelardo2015,ElGanainyMakrisKhajavikhanEtAl2018,GongAshidaKawabataEtAl2018,OkumaKawabataShiozakiEtAl2020,AshidaGongUeda2020,MatsumotoKawabataAshidaEtAl2020,DingFangMa2022,ChenSongLado2023,ZhengQiaoWangEtAl2024,ShenLuLadoEtAl2024}. 
The second application is the generalized heat equation with absorbing boundary, which is not only a fundamental model for heat processes but also useful for modeling hydrodynamical shocks in fluid dynamics, edge detection in image analysis, and option prices in finance~\cite{evans2010partial,widder1976heat,cannon1984one,thambynayagam2011diffusion}. 
For both examples, we show that quantum algorithms can solve them with fast-forwarded complexity sub-linear in time.

\subsection{Discussions and open questions}

In this work, we show that quantum algorithms can prepare history states of dissipative ODEs with cost $\Or\left( \log(T) (\log(1/\epsilon))^2 \right)$, exponentially better than previous results, and can prepare final states of dissipative ODEs with cost $\widetilde{\Or}\left( \sqrt{T} (\log(1/\epsilon))^2 \right)$, achieving a quadratic speedup in $T$. 
A natural open question is whether one can achieve an even better time scaling for final state preparation or an $\Omega(\sqrt{T})$ lower bound can be proved. 
Another open question is to generalize our fast-forwarding results to more types of ODEs beyond dissipative ones. 
One possibility might be ODEs with a time-dependent version of Lyapunov stability condition $P(t)A(t) + A(t)^{\dagger}P(t) < 0$ for an invertible time-dependent matrix $P(t)$. 
However, notice that the solution of such an ODE might still grow linearly in the worst case as we may construct a piece-wise constant $A(t)$ by gluing time-independent Lyapunov stable matrices up to their transient growing periods. 
So we expect that it is only possible for fast-forwarding under further decay assumption on $P(t)$ to rule out such growth.

In our work, we focus on analyzing the linear-system-based approach for ODEs. 
It is interesting to further explore the possibility of fast-forwarding evolution-based approach as well, especially the best linear combination of Hamiltonian simulation (LCHS) algorithm in~\cite{AnChildsLin2023}. 
We refer interested readers to our follow-up work~\cite{YangOnwuntaAn2025} for quantum dissipative ODE algorithms based on LCHS and the time-marching method~\cite{FangLinTong2022} with fast-forwarded complexity. 

Finally, we would like to remark that all the complexities in this work are query complexities, which are measured by the number of queries to the input model of the matrices (i.e., block-encoding) and the vectors (i.e., state preparation oracle). 
Although query complexities of all the algorithms do not explicitly depend on the dimension of the ODE, how the gate complexities scale in terms of the dimension depends on the efficiency of constructing the block-encoding and the state preparation oracle. 
For certain cases, these oracles can be efficiently constructed with only poly-logarithmic dimension dependence by known block-encoding techniques~\cite{GilyenSuLowEtAl2019,CladerDalzellStamatopoulosEtAl2022,NguyenKianiLloyd2022,camps2023explicit,sunderhauf2023blockencoding} and state preparation techniques~\cite{GroverRudolph2002,ZhangLiYuan2022}, and a further study on more general scenarios would be a practically important direction to fully unlock the power of our quantum dissipative ODE algorithms in a gate-based implementation. 

\subsection{Organization}

The rest of the paper is organized as follows. 
We first start with discussing the stability condition for the dissipative ODEs in~\cref{sec:stability}. 
Then, we state the quantum ODE algorithms for history state preparation and final state preparation in~\cref{sec:algorithm}, and establish the framework of obtaining fast-forwarded complexity estimates in~\cref{sec:complexity}. 
In~\cref{sec:specific_methods}, we show in a more explicit way the complexity of truncated Dyson series methods and two lower-order methods for stable dynamics. 
We discuss applications of our algorithms in~\cref{sec:applications}.

\section{Dissipative ODEs and stability}\label{sec:stability}

This work focuses on fast-forwarding quantum algorithms for dissipative ODEs. 
The dissipative ODEs in this work refers to~\cref{eqn:ODE} where the logarithmic norm of $A(t)$ is uniformly negative. 
Specifically, we assume that there exists a positive number $\eta> 0$ such that 
\begin{equation}\label{eqn:stability}
     A(t) + A(t)^{\dagger} \leq - 2 \eta < 0. 
\end{equation}
An important feature of dissipative ODEs is that the homogeneous time evolution operator always decays exponentially in time. 
We show it in the following result and give its proof in~\cref{app:proof_stability}. 

\begin{lem}\label{lem:stability_continuous_ODE}
    For any $0 \leq t_0 \leq t_1 \leq T $,  
    \begin{enumerate}
        \item we have
        \begin{equation}
            \norm{\mathcal{T} e^{\int_{t_0}^{t_1} A(s) ds} } \leq e^{-\eta (t_1-t_0)}. 
        \end{equation}
        \item for any quantum state $\ket{v}$, we have 
        \begin{equation}
         \norm{\mathcal{T} e^{\int_{t_0}^{t_1} A(s) ds} \ket{v} } \geq e^{-(t_1-t_0)\max_{t}\norm{A(t)}}. 
        \end{equation}
    \end{enumerate}
\end{lem}

Notice that here the ``dissipative'' condition is only imposed on the homogeneous time evolution operator. 
In the homogeneous case where $b(t) \equiv 0$, the solution of~\cref{eqn:ODE} can be written as $u(t) = \mathcal{T} e^{\int_{0}^{t} A(s) ds} u_0 $, so~\cref{lem:stability_continuous_ODE} implies an exponential decay of $u(t)$ in time $t$. 
However, for general inhomogeneous case with non-trivial $b(t)$, the solution of~\cref{eqn:ODE} does not necessarily decay. 
In fact, according to Duhamel's principle, the solution of~\cref{eqn:ODE} can be written as 
\begin{equation}\label{eqn:Duhamel}
    u(t) = \mathcal{T} e^{\int_{0}^{t} A(s) ds} u_0 + \int_0^t \mathcal{T} e^{\int_{\tau}^{t} A(s) ds} b(\tau) d\tau, 
\end{equation}
and the second term on the right hand side does not necessarily decay. 
Nevertheless,~\cref{eqn:Duhamel} suggests that the inhomogeneous term $b(\tau)$ at a fixed time $\tau$ only affects the dynamics locally, since the evolution operator $\mathcal{T} e^{\int_{\tau}^{t} A(s) ds}$ has an exponential decay in time.

\section{Quantum ODE algorithm}\label{sec:algorithm}

In this section, we discuss the main steps of the quantum ODE algorithm. 
Our algorithm is a unified framework of existing linear-system-based approach in~\cite{Berry2014,BerryChildsOstranderEtAl2017,ChildsLiu2020,Krovi2022}, which can be incorporated with any single-step time discretization scheme. 
We first present a general single-step time discretization approach for the ODE in~\cref{eqn:ODE}, and then discuss our algorithms for history state preparation and final state preparation.

\subsection{Time discretization}

For time discretization, let us consider a single-step scheme of the form 
\begin{equation}\label{eqn:single_step_method_homo}
    L(j,h) u_{j+1} = R(j,h) u_j + v(j,h). 
\end{equation}
Here each $u_j$ aims at approximating the solution $u(jh)$. 
$L$ and $R$ are two matrix-valued functions that might depend on the index $j$ and the step size $h$. 
$v(j,h)$ is a vector-valued function for approximating the inhomogeneous term. 
For notational convenience, we will omit the explicit $h$ dependence in the matrices and vectors and denote 
\begin{equation}
    L_j = L(j,h), \quad R_j = R(j,h), \quad v_j = v(j,h). 
\end{equation}

The simplest example of \cref{eqn:single_step_method_homo} is the forward Euler method, which is given by 
\begin{equation}
\label{FEM}
    \frac{u_{j+1} - u_j}{h} = A(jh) u_j + b(jh).
\end{equation}
Observe  that~\cref{FEM} corresponds to the form~\cref{eqn:single_step_method_homo} with $L_j = I$, $R_j = I + h A(jh)$ and $v_j = h b(jh)$. 
Another example is the trapezoidal rule
\begin{equation}
    \frac{u_{j+1} - u_j}{h} =  \frac{1}{2}(A(jh)u_j + b(jh) + A((j+1)h)u_{j+1} + b((j+1)h)), 
\end{equation}
i.e., 
\begin{equation}
    \left(I - \frac{h}{2}A((j+1)h)\right)u_{j+1} = \left(I + \frac{h}{2}A(jh)\right)u_j + \frac{h}{2}(b(jh)+b((j+1)h)).
\end{equation}
which corresponds to the form~\cref{eqn:single_step_method_homo} with
\begin{equation}
    L_j = I - \frac{h}{2}A((j+1)h), \quad
    R_j = I + \frac{h}{2}A(jh), \quad 
    v_j = \frac{h}{2}(b(jh) + b((j+1)h)). 
\end{equation}
The Dyson series method~\cite{BerryCosta2022} truncated at order $K$ can also be represented in the form of~\cref{eqn:single_step_method_homo} with $L_j = I$, 
\begin{equation}
    R_j = \sum_{k=0}^{K} \int_{jh}^{(j+1)h} d t_1  \int_{jh}^{t_1} d t_2 \cdots \int_{jh}^{t_{k-1}} d t_k A(t_1)A(t_2)\cdots A(t_k), 
\end{equation}
and 
\begin{equation}
    v_j = \sum_{k=1}^K \int_{jh}^{(j+1)h} d t_1  \int_{jh}^{t_1} d t_2 \cdots \int_{jh}^{t_{k-1}} d t_k A(t_1)A(t_2)\cdots A(t_{k-1})b(t_k).  
\end{equation}
Notice that even for a specific time discretization scheme, the choices of $L$, $R$, and $v$ are not unique, as we may always multiply both sides of~\cref{eqn:single_step_method_homo} by any matrix. 
We will choose proper $L$, $R$ and $v$ for ease of algorithmic implementation. 

\subsection{Algorithms}

\subsubsection{History state preparation}

We first aim to obtain a history state proportional to 
\begin{equation}
    \mathbf{u}_{\text{exact}} = [u(0);u(h);u(2h);\cdots;u(Mh)]
\end{equation}
encoding the information on all dynamics at $M+1$ equidistant time steps. 
The algorithm is designed based on so-called the all-at-once system, which is a dilated linear system of equations 
\begin{equation}
    \mathbf{A}_{M,0} \mathbf{u} = \mathbf{b}_{M,0}. 
\end{equation} 
Here 
\begin{equation}\label{eqn:def_A_b}
    \mathbf{A}_{M,0} = \left( \begin{array}{ccccc}
        I & & & & \\
        -R_0 & L_0 & & & \\
         & -R_1 & L_1 & & \\
         & & \ddots &\ddots & \\
         & & & -R_{M-1} & L_{M-1} 
    \end{array} \right), \quad 
    \mathbf{b}_{M,0} = \left( \begin{array}{c}
        u_0 \\
        v_0 \\
        v_1 \\
        \vdots \\
        v_{M-1}
    \end{array} \right). 
\end{equation}
We add two subscripts for $\mathbf{A}$ and $\mathbf{b}$. 
The first subscript indicates the overall number of time steps in the discrete evolution. 
The second subscript denotes the number of padding rows, whose meaning will be clear later. 
In the case of history state preparation, the second subscript is always $0$.

A quantum algorithm for estimating a history state proportional to $\mathbf{u}_{\text{exact}} = [u(0);u(h);u(2h);\cdots;u(T)]$ can then be constructed as follows. 
We first construct a state preparation oracle of $\mathbf{b}_{M,0}$ and a block-encoding of $\mathbf{A}_{M,0}$, defined in~\cref{eqn:def_A_b}.  
The state preparation oracle of $\mathbf{b}_{M,0}$ requires access to $\ket{u_0}$ and $b(s)$, and the block-encoding of $\mathbf{A}_{M,0}$ requires access to $A(t)$. 
Constructions of both oracles typically depend on specific numerical methods, and we will discuss some examples in~\cref{sec:specific_methods}. 
Then, with these input models of $\mathbf{A}_{M,0}$ and $\mathbf{b}_{M,0}$, we can apply the optimal quantum linear system algorithm~\cite{CostaAnYuvalEtAl2022,Dalzell2024} to approximate $\ket{\mathbf{u}} = \mathbf{A}_{M,0}^{-1} \mathbf{b}_{M,0} / \norm{ \mathbf{A}_{M,0}^{-1} \mathbf{b}_{M,0} } $, which is the desired approximation of the history state.

\subsubsection{Final state preparation}

Now we discuss quantum algorithms for preparing the state $u(T)/\norm{u(T)}$ only encoding the solution at the final time. 
A common strategy to obtain $u(T)/\norm{u(T)}$ is to post select the history state $[u(0);u(h);\cdots;u(T)]$ obtained by the history state preparation algorithm. 
Notice that the history state can be written as (up to a normalization factor) $\sum_{j=0}^M \norm{u(jh)} \ket{j}\ket{u(jh)}$, so if we measure the first ancilla register onto $\ket{M}$, then the system register will encode the desired final solution. 
However, the probability of getting $M$ can be small as we typically use many time steps for accurate simulation. 
To boost such probability, we may use the padding trick, introduced in~\cite{Berry2014}, to make multiple copies of the final solution in the history state. 

Specifically, for a single step method, instead of solving the linear system defined by~\cref{eqn:def_A_b}, now we consider $\mathbf{A}_{M,M_p-1} \mathbf{u} = \mathbf{b}_{M,M_p-1} $ where 
\begin{align}
    \mathbf{A}_{M,M_p-1} &= \left( \begin{array}{ccccccccc}
        I & & & & & & & & \\
        -R_0 & L_0 & & & & & & &  \\
         & -R_1 & L_1 & & & & & &  \\
         & & \ddots &\ddots & & & & &  \\
         & & & -R_{M-1} & L_{M-1} & & & & \\
         & & & & -I & I & & & \\
         & & & &  & -I & I & & \\ 
         & & & &  &  & \ddots & \ddots & \\ 
         & & & &  & & & -I & I \\ 
    \end{array} \right), \label{eqn:def_A_padding}\\
    \mathbf{b}_{M,M_p-1} &= [ u_0; v_0; v_1; \cdots; v_{M-1}; 0; 0; \cdots; 0]. \label{eqn:def_b_padding}
\end{align}
Here $\mathbf{A}_{M,M_p-1}$ is a dilation of that in~\cref{eqn:def_A_b} by having $(M_p-1)$ many additional rows, of which each has an identity on the diagonal block and a negative identity on the sub-diagonal block. 
$\mathbf{b}_{M,M_p-1}$ is correspondingly dilated by appending zeros. 
The second subscript $M_p - 1$ denotes the number of the additional rows, where $M_p$ is a positive integer. 
Notice that the case $M_p = 1$ implies no padding row and becomes the matrix~\cref{eqn:def_A_b} used in history state preparation. 
Then, the exact solution of the linear system $\mathbf{A}_{M,M_p-1} \mathbf{u} = \mathbf{b}_{M,M_p-1} $ becomes $\mathbf{u} = [u_0;u_1;u_2;\cdots;u_{M-1};u_M;u_M;\cdots;u_M]$, which is the history solution (up to step $M-1$) followed by $M_p$ copies of the final solution $u_M$. 

Our quantum algorithm for preparing the final solution first solves the linear system defined by~\cref{eqn:def_A_padding} and~\cref{eqn:def_b_padding} using optimal quantum linear system algorithm~\cite{CostaAnYuvalEtAl2022,Dalzell2024}, then measures the index register until success. 
The algorithm succeeds if the measurement outcome is larger than or equal to $M$. 
To further reduce the complexity quadratically, we may combine it with the amplitude amplification technique~\cite{BrassardHoyerMoscaEtAl2002}.

The choice of $M_p$ heavily affects the complexity of the algorithm. 
If $M_p$ is too small, then there will be too few copies of the final solution and thus small success probability. 
If $M_p$ is too large, then the condition number of $\mathbf{A}_{M,M_p-1}$ will increase, and solving corresponding linear system of equations becomes expensive. 
Later we will present a complexity estimate for any value of $M_p$ and any single-step method, and discuss optimal choices of $M_p$ in specific schemes.

\section{Fast-forwarded complexity}\label{sec:complexity}

Here we establish our fast-forwarded complexity analysis for dissipative ODEs. 
First, for any value of $M$ and $M_p$, we bound the condition number of the matrix $\mathbf{A}_{M,M_p-1}$ in the linear system~\cref{eqn:def_A_padding}, which is the dominant factor in quantum linear system algorithms. 
Then we estimate the complexity of history state preparation, and final state preparation separately. 
For history state preparation, we give a complexity estimate for general inhomogeneous ODEs. 
In the special case of homogeneous ODEs, although the general theorem also applies, by more careful analysis we may obtain a better complexity, which we will present separately. 
For final state preparation, we will only focus on the general inhomogeneous case, as the final solution of the homogeneous dissipative ODE always decays exponentially and becomes trivial. 

\subsection{Condition number}

We estimate the condition number of $\mathbf{A}_{M,M_p-1}$ by bounding $\norm{\mathbf{A}_{M,M_p-1}}$ and $\norm{\mathbf{A}_{M,M_p-1}^{-1}}$. Our main technique is a bound of the spectral norm of a block matrix through the spectral norms of its blocks, which we state here and prove in~\cref{app:proof_2norm_block_bound}. 

\begin{lem}\label{lem:2_norm_block_bound}
    Let $\mathbf{M}$ be the block matrix as 
    \begin{equation}
        \mathbf{M} = \left(\begin{array}{ccc}
            M_{0,0} & \cdots & M_{0,n-1} \\
            \vdots & & \vdots \\
            M_{n-1,0} & \cdots & M_{n-1,n-1}
        \end{array}\right)
    \end{equation}
    where $M_{ij}$'s are $d$ dimensional square matrices. 
    Then 
    \begin{equation}
        \|\mathbf{M}\| \leq \sqrt{ \left( \max_j  \sum_{k=0}^{n-1}  \norm{M_{k,j}}\right) \left( \max_i  \sum_{k=0}^{n-1} \norm{M_{i,k}} \right)  }. 
    \end{equation}
\end{lem}

The spectral norm of $\mathbf{A}_{M,M_p-1}$ can be bounded straightforwardly from~\cref{lem:2_norm_block_bound}. 
For $\mathbf{A}_{M,M_p-1}^{-1}$, we may first perform block version of Gaussian elimination to have an explicit form of $\mathbf{A}_{M,M_p-1}^{-1}$, then apply~\cref{lem:2_norm_block_bound} to bound its spectral norm.

\begin{lem}\label{lemma:condition-number}
    Let $A(t)$ be a matrix such that $A(t) + A(t)^{\dagger} \leq -2\eta < 0$, and $\mathbf{A}_{M,M_p-1}$ be the matrix defined in~\cref{eqn:def_A_padding}. 
    Suppose that $\eta h \leq 1$ and the local truncation error is bounded as 
    \begin{equation}\label{eqn:num_assump}
        \norm{L_j^{-1} R_j - \mathcal{T} e^{\int_{jh}^{(j+1)h} A(s) ds}  } \leq \frac{1}{2}\eta h e^{-\eta h}, \quad \forall 0 \leq j \leq M-1. 
    \end{equation}
    Then, for any $M > T$, we have 
    \begin{align}
        \norm{\mathbf{A}_{M,M_p-1}} &\leq 2 + \max_j \norm{L_j} + \max_j \norm{R_j}, \\
        \norm{\mathbf{A}_{M,M_p-1}^{-1}} &\leq \left( \frac{2eM}{\eta T} + M_p \right) \left( 1+\max_j \norm{L_j^{-1}} \right). 
    \end{align}
\end{lem}
\begin{proof}
    The upper bound of $\norm{\mathbf{A}_{M,M_p-1}}$ can be directly obtained by applying~\cref{lem:2_norm_block_bound} and noticing that the block row sums and column sums of $\norm{\mathbf{A}_{M,M_p-1}}$ are both bounded by $2 + \max_j \norm{L_j} + \max_j \norm{R_j}$. 

    We focus on bounding $\norm{\mathbf{A}_{M,M_p-1}^{-1}}$. 
   For notation simplicity, let us denote 
   \begin{equation}
       L_{j} = I, \quad j = -1 \text{ or }M \leq j \leq M+M_p-2,
   \end{equation}
   and 
   \begin{equation}
       R_{j} = I, \quad M \leq j \leq M+M_p-2. 
   \end{equation}
    Then we can simply write 
    \begin{align}
    \mathbf{A}_{M,M_p-1} &= \left( \begin{array}{ccccc}
        L_{-1} & & & & \\
        -R_0 & L_0 & & & \\
         & -R_1 & L_1 & & \\
         & & \ddots & \ddots & \\
         & & & -R_{M+M_p-2} & L_{M+M_p-2}
    \end{array} \right). 
\end{align}
    By Gaussian elimination, we have $\mathbf{A}_{M,M_p-1}^{-1} = \left( B_{i,j} \right)$, where the block
    \begin{equation}
        B_{i,j} = \mathbf{1}_{i\geq j} \left( \prod_{l=j}^{i-1} L_l^{-1}R_l \right)L_{j-1}^{-1}. 
    \end{equation}

    To apply~\cref{lem:2_norm_block_bound}, we first bound the spectral norm of the block $B_{i,j}$. 
    Notice that, for any $0 \leq j \leq M-1$, according to~\cref{eqn:num_assump} and~\cref{lem:stability_continuous_ODE}, we have 
    \begin{align}
        \norm{L_j^{-1} R_j } \leq \norm{ L_j^{-1} R_j - \mathcal{T} e^{\int_{jh}^{(j+1)h} A(s) ds}  } + \norm{\mathcal{T} e^{\int_{jh}^{(j+1)h} A(s) ds} }  \leq \frac{1}{2}\eta h e^{-\eta h} + e^{-\eta h} \leq e^{-\eta h/2}, 
    \end{align}
    where the last inequality is because $1+x\leq e^x$ for any $x \geq 0$. 
    In other words, we may write 
    \begin{equation}\label{eqn:num_stable_assump}
        \norm{L_j^{-1} R_j } \leq c^h, \quad 0 \leq j \leq M-1, 
    \end{equation}
    for a constant 
    \begin{equation}\label{eqn:proof_cond_num_def_c}
        c = e^{-\eta/2}. 
    \end{equation}
    Notice that $L_j^{-1} = R_j = I$ for $j \geq M$. 
    Then 
    \begin{equation}
        \norm{B_{i,j}} \leq \mathbf{1}_{i\geq j} \left( \prod_{l=j}^{i-1} \norm{L_l^{-1}R_l} \right) \left( 1+\max_l \norm{L_l^{-1}} \right) \leq \begin{cases}
            c^{(i-j)h} \left( 1+\max_l \norm{L_l^{-1}} \right) , & M-1 \geq i \geq j, \\
            c^{(M-j)h}  \left( 1+\max_l \norm{L_l^{-1}} \right) , & i \geq M > j, \\
             1+\max_l \norm{L_l^{-1}} , & i \geq j \geq M, \\
            0, & i < j. 
        \end{cases}
    \end{equation}
    Using this upper bound of $\norm{B_{i,j}}$, we may bound the block row sums of $\mathbf{A}_{M,M_p-1}^{-1}$ as 
    \begin{align}
        \sum_{k=0}^{M+M_p-1} \norm{B_{i,k}} &\leq \begin{cases}
            \sum_{k=0}^{i} c^{(i-k)h} \left( 1+\max_l \norm{L_l^{-1}} \right) ,& i \leq M - 1, \\
             \left(\sum_{k=0}^{M} c^{(M-k)h} + i-M\right) \left( 1+\max_l \norm{L_l^{-1}} \right) ,& i \geq M 
        \end{cases}  \\
        & \leq \begin{cases}
            \frac{1}{1-c^{h}} \left( 1+\max_l \norm{L_l^{-1}} \right) ,& i \leq M - 1, \\
             \left(\frac{1}{1-c^{h}} + i-M\right) \left( 1+\max_l \norm{L_l^{-1}} \right) ,& i \geq M 
        \end{cases} \\
        & \leq \left(\frac{1}{1-c^{h}} + M_p \right) \left( 1+\max_l \norm{L_l^{-1}} \right), 
    \end{align}
    and the block column sums as 
    \begin{align}
        \sum_{k=0}^{M+M_p-1} \norm{B_{k,j}} &\leq \begin{cases}
            \left( \sum_{k=j}^{M-1} c^{(k-j)h} + M_p c^{(M-j)h} \right) \left( 1+\max_l \norm{L_l^{-1}} \right) ,& j \leq M - 1, \\
             (M+M_p-j) \left( 1+\max_l \norm{L_l^{-1}} \right) ,& j \geq M 
        \end{cases} \\
        & \leq \begin{cases}
            \left( \frac{1}{1-c^h} + M_p \right) \left( 1+\max_l \norm{L_l^{-1}} \right) ,& j \leq M - 1, \\
             (M+M_p-j) \left( 1+\max_l \norm{L_l^{-1}} \right) ,& j \geq M 
        \end{cases} \\
        & \leq \left( \frac{1}{1-c^h} + M_p \right) \left( 1+\max_l \norm{L_l^{-1}} \right). 
    \end{align}
    According to~\cref{lem:2_norm_block_bound}, we have 
    \begin{equation}
        \norm{\mathbf{A}_{M,M_p-1}^{-1}} \leq \left( \frac{1}{1-c^h} + M_p \right) \left( 1+\max_l \norm{L_l^{-1}} \right). 
    \end{equation}
     Notice that $1-c^{h} = 1- e^{-\eta h/2} \geq \eta h/(2e) $ since the inequality $1 - e^{x} \geq -x/e$ holds for $x \in [-1,0]$. 
     Thus,
     \begin{equation}
         \norm{\mathbf{A}_{M,M_p-1}^{-1}} \leq \left( \frac{2e}{\eta h} + M_p \right) \left( 1+\max_l \norm{L_l^{-1}} \right) = \left( \frac{2e}{\eta} \frac{M}{T} + M_p \right) \left( 1+\max_l \norm{L_l^{-1}} \right). 
     \end{equation}
\end{proof}

\subsection{Complexity of history state preparation}

In the history state preparation, we choose $M_p = 1$. 
We first derive quantum complexity in the general inhomogeneous case, then present a simpler version only applicable to the homogeneous case. 

\subsubsection{General inhomogeneous case}

\begin{thm}\label{thm:complexity_single_step_inhomo}
    Consider the ODE in~\cref{eqn:ODE} such that $A(t) + A(t)^{\dagger} \leq -2\eta < 0$. 
    Let $\epsilon > 0$ be the target error, $T \geq (\max\norm{A(t)})^{-1}$ be the evolution time, and $M$ be the number of time steps. 
    Suppose that we use the single-step method in~\cref{eqn:single_step_method_homo} for its time discretization, the time step size $h = T/M$ is chosen such that $\eta h \leq 1$ and for all $0 \leq j \leq T/h-1$, 
    \begin{equation}\label{eqn:num_assump_complexity_inhomo_1}
        \norm{L_j^{-1} R_j - \mathcal{T} e^{\int_{jh}^{(j+1)h} A(s) ds}  } \leq \min\left\{ \frac{1}{2} \eta h e^{-\eta h}, \frac{ \eta^{3/2} h \epsilon }{ 144\sqrt{2} \sqrt{1+\frac{T\max\norm{b(t)}^2 }{\eta \norm{u_0}^2 }}  \sqrt{\max\norm{A(t)} + \frac{\max\norm{b(t)}}{\norm{u_0}} } } \right\}, 
    \end{equation}
    and 
    \begin{equation}\label{eqn:num_assump_complexity_inhomo_2}
        \norm{ L_j^{-1} v_j - \int_{jh}^{(j+1)h} \mathcal{T} e^{\int_{s}^{(j+1)h} A(\tau) d\tau} b(s) ds  } \leq \frac{\norm{u_0} \eta h \epsilon}{ 72\sqrt{2} \sqrt{T} \sqrt{ \max\norm{A(t)} + \frac{\max\norm{b(t)}}{\norm{u_0}} } } . 
    \end{equation}
    Then, we can obtain an $\epsilon$-approximation of the history state $\mathbf{u}_{\text{exact}}/\norm{\mathbf{u}_{\text{exact}}}$, using 
    \begin{equation}
        \Or\left( \left( 1 + \max_j \norm{L_j} + \max_j \norm{R_j} \right) \left( 1+\max_j \norm{L_j^{-1}} \right)  \frac{M}{\eta T} \log\left(\frac{1}{\epsilon}\right) \right)
    \end{equation}
    queries to the block-encoding of $\mathbf{A}_{M,0}$ and the state preparation oracle of $\ket{\mathbf{b}_{M,0}}$. 
\end{thm}
\begin{proof}
    We first show that, under the assumptions in the statement of the theorem, the exact normalized solution of the linear system $\mathbf{A}_{M,0} \mathbf{u} = \mathbf{b}_{M,0}$ is an $\epsilon/2$-approximation of the ideal history state. 
    
    Let $M = T/h$, $\mathbf{u} = \mathbf{A}_{M,0}^{-1} \mathbf{b}_{M,0}$ and $\ket{\mathbf{u}} = \mathbf{u} / \norm{\mathbf{u}}$. 
    We denote $L_{-1} = I$. 
    Let 
    \begin{equation}\label{eqn:proof_complexity_inhomo_epsilon_def}
        \epsilon_1 = \frac{ \eta^{3/2} h \epsilon }{ 144\sqrt{2} \sqrt{1+\frac{T\max\norm{b}^2 }{\eta \norm{u_0}^2 }}  \sqrt{\max\norm{A(t)} + \frac{\max\norm{b(t)}}{\norm{u_0}} } }, \quad \epsilon_2 = \frac{\norm{u_0} \eta h \epsilon}{ 72\sqrt{2} \sqrt{T} \sqrt{ \max\norm{A(t)} + \frac{\max\norm{b(t)}}{\norm{u_0}}  } }. 
    \end{equation}
    Then the assumptions on the local operators become  
    \begin{equation}
        \norm{L_j^{-1} R_j - \mathcal{T} e^{\int_{jh}^{(j+1)h} A(s) ds}  } \leq \epsilon_1 , 
    \end{equation}
    and 
    \begin{equation}
        \norm{ L_j^{-1} v_j - \int_{jh}^{(j+1)h} \mathcal{T} e^{\int_{s}^{(j+1)h} A(\tau) d\tau} b(s) ds  } \leq \epsilon_2. 
    \end{equation}
    The numerical scheme becomes 
    \begin{equation}\label{eqn:proof_complexity_inhomo_num_scheme}
        u_{j+1} = L_j^{-1} R_j u_j + L_j^{-1} v_j = P_j u_j + L_j^{-1} v_j,
    \end{equation}
    where we denote 
    \begin{equation}
        P_j = L_j^{-1} R_j. 
    \end{equation}
    Thus, applying~\cref{eqn:proof_complexity_inhomo_num_scheme} iteratively, we have 
    \begin{equation}\label{eqn:proof_complexity_inhomo_num_solu}
        u_k = \left( \prod_{j=0}^{k-1} P_j \right) u_0 + \sum_{j=0}^{k-1} \left( \prod_{l=j+1}^{k-1} P_l \right) L_j^{-1} v_j. 
    \end{equation}
    Similarly, for the exact solution $u(t)$, by Duhamel's principle we have 
    \begin{equation}
        u((j+1)h) = \mathcal{T} e^{\int_{jh}^{(j+1)h} A(s) ds } u(jh) + \int_{jh}^{(j+1)h} \mathcal{T} e^{\int_{s}^{(j+1)h} A(\tau) d\tau } b(s) ds = Q_j u(jh) + w_j, 
    \end{equation}
    where we denote 
    \begin{equation}
        Q_j = \mathcal{T} e^{\int_{jh}^{(j+1)h} A(s) ds }, \quad w_j = \int_{jh}^{(j+1)h} \mathcal{T} e^{\int_{s}^{(j+1)h} A(\tau) d\tau } b(s) ds. 
    \end{equation}
    Hence, we have 
    \begin{equation}\label{eqn:proof_complexity_inhomo_exact_solu}
        u(kh) = \left( \prod_{j=0}^{k-1} Q_j \right) u_0 + \sum_{j=0}^{k-1} \left( \prod_{l=j+1}^{k-1} Q_l \right) w_j. 
    \end{equation}
    Subtracting~\cref{eqn:proof_complexity_inhomo_exact_solu} from~\cref{eqn:proof_complexity_inhomo_num_solu} and using triangle inequality, we have 
    \begin{equation}\label{eqn:proof_complexity_inhomo_bound_1}
        \norm{u_k - u(kh)} \leq \norm{ \prod_{j=0}^{k-1} P_j - \prod_{j=0}^{k-1} Q_j } \norm{u_0} + \sum_{j=0}^{k-1} \norm{ \left( \prod_{l=j+1}^{k-1} P_l \right) L_j^{-1} v_j - \left( \prod_{l=j+1}^{k-1} Q_l \right) w_j }. 
    \end{equation}

    For the first term in the right-hand side of~\cref{eqn:proof_complexity_inhomo_exact_solu}, we use the triangle inequality to write 
    \begin{align}
        \norm{  \prod_{j=0}^{k-1} P_j - \prod_{j=0}^{k-1} Q_j  } & \leq \norm{ \sum_{l=0}^{k-1} \left[ \left(\prod_{j=l+1}^{k-1}P_j\right) (P_l - Q_l) \left(\prod_{j=0}^{l-1} Q_j\right) \right] } \\
        & \leq \sum_{l=0}^{k-1} \left[ \left(\prod_{j=l+1}^{k-1}\norm{P_j}\right) \norm{P_l - Q_l} \left(\prod_{j=0}^{l-1} \norm{Q_j} \right) \right] \\
        & \leq \epsilon_1 \sum_{l=0}^{k-1} \left(\prod_{j=l+1}^{k-1}\norm{P_j}\right) \left(\prod_{j=0}^{l-1} \norm{Q_j} \right). 
    \end{align}
    By~\cref{lem:stability_continuous_ODE} and the proof of~\cref{lemma:condition-number}, we have $\norm{Q_j} \leq e^{-\eta h} \leq e^{-\eta h/2}$ and $\norm{P_j} \leq e^{-\eta h/2}$, and thus 
    \begin{equation}\label{eqn:proof_complexity_inhomo_bound_2}
        \norm{  \prod_{j=0}^{k-1} P_j - \prod_{j=0}^{k-1} Q_j  } \leq k \epsilon_1 e^{-\eta (k-1)h/2}. 
    \end{equation}
    For the second term in the right-hand side of~\cref{eqn:proof_complexity_inhomo_bound_1}, we again use the triangle inequality to obtain (the following derivation assumes $j \leq k - 2$, but the final conclusion also holds true for $j = k-1$)
    \begin{align}
        & \quad \norm{ \left( \prod_{l=j+1}^{k-1} P_l \right) L_j^{-1} v_j - \left( \prod_{l=j+1}^{k-1} Q_l \right) w_j } \\
        & \leq \norm{ \left( \prod_{l=j+1}^{k-1} P_l - \prod_{l=j+1}^{k-1} Q_l \right)  w_j } + \norm{ \left(\prod_{l=j+1}^{k-1} P_l\right) \left( L_j^{-1} v_j - w_j \right) } \\
        & \leq \norm{\prod_{l=j+1}^{k-1} P_l - \prod_{l=j+1}^{k-1} Q_l}\norm{w_j} +  \left(\prod_{l=j+1}^{k-1} \norm{P_l}\right) \norm{L_j^{-1} v_j - w_j}. 
    \end{align}
    Using $\norm{P_l} \leq e^{-\eta h/2}$, $\norm{w_j} \leq h \max\norm{b(t)}$ from the definition of $w_j$, and $\norm{\prod_{l=j+1}^{k-1} P_l - \prod_{l=j+1}^{k-1} Q_l} \leq (k-j-1)\epsilon_1 e^{-\eta (k-j-2)h/2} $ which can be obtained by the same proof of~\cref{eqn:proof_complexity_inhomo_bound_2}, we have 
    \begin{equation}\label{eqn:proof_complexity_inhomo_bound_3}
        \norm{ \left( \prod_{l=j+1}^{k-1} P_l \right) L_j^{-1} v_j - \left( \prod_{l=j+1}^{k-1} Q_l \right) w_j } \leq (k-j-1)\epsilon_1 e^{-\eta (k-j-2)h/2} h \max\norm{b(t)} + e^{-\eta (k-j-1)h/2} \epsilon_2. 
    \end{equation}

    Plugging~\cref{eqn:proof_complexity_inhomo_bound_2,eqn:proof_complexity_inhomo_bound_3} back to~\cref{eqn:proof_complexity_inhomo_bound_1}, we have 
    \begin{align}
         & \quad \norm{u_k - u(kh)} \\
         & \leq  k \epsilon_1 e^{-\eta (k-1)h/2}\norm{u_0} + \sum_{j=0}^{k-1} (k-j-1)\epsilon_1 e^{-\eta (k-j-2)h/2} h \max\norm{b(t)} + \sum_{j=0}^{k-1} e^{-\eta (k-j-1)h/2} \epsilon_2 \\
         & = k \epsilon_1 e^{-\eta (k-1)h/2}\norm{u_0} + \epsilon_1 h \max\norm{b(t)} \frac{ 1-e^{-\eta k h/2} - k e^{-\eta (k-1)h/2} (1-e^{-\eta h/2})  }{(1-e^{-\eta h/2})^2} + \epsilon_2 \frac{1-e^{-\eta k h/2}}{1-e^{-\eta h/2}} \\
         & \leq k \epsilon_1 e^{-\eta (k-1)h/2}\norm{u_0} + \frac{\epsilon_1 h \max\norm{b(t)}}{(1-e^{-\eta h/2})^2} + \frac{\epsilon_2}{1-e^{-\eta h/2}}. 
    \end{align}
    Then the error in the unnormalized history vector can be bounded as 
    \begin{align}
        \norm{\mathbf{u} - \mathbf{u}_{\text{exact}}}
        &= \sqrt{ \sum_{k=1}^{M} \norm{ u_k - u(kh) }^2  } \\
        & \leq \sqrt{3} \sqrt{ \sum_{k=1}^M k^2 \epsilon_1^2 e^{-\eta (k-1)h} \norm{u_0}^2 + \sum_{k=1}^M \frac{\epsilon_1^2 h^2 \max\norm{b(t)}^2}{(1-e^{-\eta h/2})^4} + \sum_{k=1}^M \frac{\epsilon_2^2 }{(1-e^{-\eta h/2})^2 } } \\
        & = \sqrt{3} \sqrt{ \sum_{k=1}^M k^2 \epsilon_1^2 e^{-\eta (k-1)h} \norm{u_0}^2 +  \frac{\epsilon_1^2 T h \max\norm{b(t)}^2}{(1-e^{-\eta h/2})^4} + \frac{\epsilon_2^2 (T/h) }{(1-e^{-\eta h/2})^2 } }. 
    \end{align}
    Notice that 
    \begin{align}
        \sum_{k=1}^M k^2 e^{-\eta (k-1)h} 
        & = \frac{(2M^2 + 2M -1)e^{-\eta(M+1)h} - M^2 e^{-\eta(M+2)h} - (M+1)^2e^{-\eta Mh} + e^{-\eta h} + 1}{(1-e^{-\eta h})^3} \\
        & =  \frac{1+e^{-\eta h}}{(1-e^{-\eta h})^3} - \frac{M^2 e^{-\eta Mh} ( 1- e^{-\eta h } )^2 }{(1-e^{-\eta h})^3} -  \frac{ 2Me^{-\eta Mh} (1-e^{-\eta h}) }{(1-e^{-\eta h})^3} - \frac{e^{-\eta M h}(1+e^{-\eta h}) }{(1-e^{-\eta h})^3}  \\
        & \leq \frac{2}{(1-e^{-\eta h})^3}, 
    \end{align}
    so, together with the inequality $1-e^{-x} \geq x/2$ for $x \in [0,1]$, we have 
    \begin{align}
        \norm{\mathbf{u} - \mathbf{u}_{\text{exact}}} 
        & \leq \sqrt{3} \sqrt{ \frac{2 \epsilon_1^2 \norm{u_0}^2 }{(1-e^{-\eta h})^3} + \frac{\epsilon_1^2 T h \max\norm{b(t)}^2}{(1-e^{-\eta h/2})^4} + \frac{\epsilon_2^2 (T/h) }{(1-e^{-\eta h/2})^2 }  } \\
        & \leq \sqrt{ \frac{48 \epsilon_1^2 \norm{u_0}^2 }{\eta^3 h^3 } + \frac{ 48 \epsilon_1^2 T \max\norm{b(t)}^2}{\eta^4 h^3 } + \frac{ 12 \epsilon_2^2 T }{\eta^2 h^3 } } . \label{eqn:proof_complexity_inhomo_bound_4}
    \end{align}

    The error in the (normalized) quantum states is bounded as 
    \begin{equation}\label{eqn:proof_complexity_inhomo_bound_6}
         \norm{\ket{\mathbf{u}} - \ket{\mathbf{u}_{\text{exact}}} } \leq \frac{2}{\norm{\mathbf{u}_{\text{exact}}} } \norm{\mathbf{u} - \mathbf{u}_{\text{exact}}}, 
    \end{equation}
    so we need to find a lower bound of $\norm{\mathbf{u}_{\text{exact}}}$. 
    To this end, we will find a lower bound of the number of $u(kh)$ such that its norm is bounded from below by a constant. 
    Specifically, by Duhamel's principle and~\cref{lem:stability_continuous_ODE}, we have 
    \begin{align}
        \norm{u(kh)} &= \norm{ \mathcal{T} e^{\int_0^{kh} A(s) ds} u_0 + \int_0^{kh} \mathcal{T} e^{\int_s^{kh} A(\tau) d\tau } b(s) ds } \\
        & \geq \norm{ \mathcal{T} e^{\int_0^{kh} A(s) ds} u_0 } - \norm{ \int_0^{kh} \mathcal{T} e^{\int_s^{kh} A(\tau) d\tau } b(s) ds } \\
        & \geq e^{- kh \max\norm{A(t)}} \norm{u_0} - kh \max\norm{b(t)}. 
    \end{align}
    Notice that for $k \leq \frac{1}{3h\max\left\{\max\norm{A(t)},\max \norm{b(t)}/\norm{u_0}\right\}}$, we always have $ kh \leq  \frac{\log(3/2)}{\max\norm{A(t)}} $ and $kh \leq \frac{\norm{u_0}}{3\max\norm{b(t)}} $, which imply $e^{- kh \max\norm{A(t)}} \geq 2/3$ and $kh \max\norm{b(t)} \leq \norm{u_0}/3$, and thus $\norm{u(kh)} \geq \norm{u_0}/3$. 
    Then 
    \begin{align}
        \norm{\mathbf{u}_{\text{exact}}} = \sqrt{ \sum_{k=0}^M \norm{u(kh)}^2 } &\geq \sqrt{ \frac{\norm{u_0}^2 }{9} \frac{1}{3h\max\left\{\max\norm{A(t)},\max \norm{b(t)}/\norm{u_0}\right\} } } \\
        & = \frac{\norm{u_0}}{3\sqrt{3}} \frac{1}{\sqrt{h \max\left\{\max\norm{A(t)},\max \norm{b(t)}/\norm{u_0}\right\} } }. \label{eqn:proof_complexity_inhomo_bound_5}
    \end{align}
    Plugging~\cref{eqn:proof_complexity_inhomo_bound_4,eqn:proof_complexity_inhomo_bound_5} back to~\cref{eqn:proof_complexity_inhomo_bound_6}, we have 
    \begin{equation}
        \norm{\ket{\mathbf{u}} - \ket{\mathbf{u}_{\text{exact}}} } \leq 36 \sqrt{\max\left\{\max\norm{A(t)},\max \frac{\norm{b(t)}}{\norm{u_0}}\right\}} \sqrt{ \frac{4 \epsilon_1^2 }{\eta^3 h^2 } + \frac{ 4 \epsilon_1^2 T \max\norm{b(t)}^2}{\eta^4 h^2 \norm{u_0}^2 } + \frac{ \epsilon_2^2 T }{\eta^2 h^2 \norm{u_0}^2 } }, 
    \end{equation}
    which is further upper bounded by $\epsilon/2$ by the choices of $\epsilon_1$ and $\epsilon_2$ in~\cref{eqn:proof_complexity_inhomo_epsilon_def}. 

    Then, solving the linear system of equations $\mathbf{A}_{M,0} \mathbf{u} = \mathbf{b}_{M,0}$ up to error $\epsilon/2$ gives the $\epsilon$-approximation of $\ket{\mathbf{u}_{\text{exact}}}$. 
    According to~\cite{CostaAnYuvalEtAl2022}, it requires $\Or( \kappa \log(1/\epsilon) )$ queries to the block-encoding of $\mathbf{A}_{M,0}$ and the state preparation oracle of $\mathbf{b}_{M,0}$, where $\kappa = \norm{\mathbf{A}_{M,0}} \norm{\mathbf{A}_{M,0}^{-1}}$ is the condition number of $\mathbf{A}_{M,0}$. 
    According to~\cref{lemma:condition-number}, we have 
    \begin{equation}
        \kappa = \Or\left( \left( 1 + \max_j \norm{L_j} + \max_j \norm{R_j} \right) \left( 1+\max_j \norm{L_j^{-1}} \right)  \frac{M}{\eta T}  \right)
    \end{equation}
    and complete the proof. 
\end{proof}

\subsubsection{Homogeneous case}

Now we consider the special homogeneous case where $b(t) \equiv 0$. 
Notice that~\cref{thm:complexity_single_step_inhomo} directly applies to the homogeneous case by letting $b(t) = 0$ in the complexity. 
However, we can obtain a better homogeneous complexity estimate by leveraging tighter discretization error bound. 
We show it in the next result and give its proof in~\cref{app:proof_complexity_single_step_homo}.

\begin{thm}\label{thm:complexity_single_step_homo}
    Consider the ODE in~\cref{eqn:ODE} with $b(t) \equiv 0$ and $A(t) + A(t)^{\dagger} \leq -2\eta < 0$. 
    Let $\epsilon > 0$ be the target error, $T \geq (\max\norm{A(t)})^{-1}$ be the evolution time, and $M$ be the number of time steps. 
    Suppose we use the single step method in~\cref{eqn:single_step_method_homo} for its time discretization, the time step size $h = T/M$ is chosen such that $\eta h \leq 1$ and 
    \begin{equation}\label{eqn:num_assump_complexity_homo}
        \norm{L_j^{-1} R_j - \mathcal{T} e^{\int_{jh}^{(j+1)h} A(s) ds}  } \leq \min \left\{ \frac{1}{2}\eta h e^{-\eta h}, \frac{ \eta^{3/2} h \epsilon }{ 32 \sqrt{  \max \norm{A(t)}} } \right\}, \quad  0 \leq j \leq T/h-1.   
    \end{equation}
    Then, we can obtain an $\epsilon$-approximation of the history state $\mathbf{u}_{\text{exact}}/\norm{\mathbf{u}_{\text{exact}}}$, using 
    \begin{equation}
        \Or \left( \left( 1 + \max_j \norm{L_j} + \max_j \norm{R_j} \right) \left( 1+\max_j \norm{L_j^{-1}} \right)  \frac{M}{\eta T} \log\left(\frac{1}{\epsilon}\right) \right)
    \end{equation}
    queries to the block-encoding of $\mathbf{A}_{M,0}$ and the state preparation oracle of $\ket{u_0}$. 
\end{thm}

\subsection{Complexity of final state preparation}

For the final state preparation, there are two components of the complexity: the cost of solving linear system of equations $\mathbf{A}_{M,M_p-1} \mathbf{u} = \mathbf{b}_{M,M_p-1}$, and the probability of measuring the ancilla register onto the correct index. 
As we discussed earlier, the choice of $M_p$ heavily affects the overall complexity. 
Here we first present a complexity estimate for any choice of $M_p$.

\begin{thm}\label{thm:complexity_final_state}
    Consider the ODE in~\cref{eqn:ODE} such that $A(t) + A(t)^{\dagger} \leq -2\eta < 0$. 
    Let $\epsilon > 0$ be the target error, $T \geq (\max\norm{A(t)})^{-1}$ be the evolution time, $M$ be the number of time steps, and $M_p$ be the number of copies of the final state in the linear system of equations. 
    Suppose we use the single step method in~\cref{eqn:single_step_method_homo} for its time discretization, the time step size $h = T/M$ is chosen such that $\eta h \leq 1$ and for all $0 \leq j \leq T/h-1$, 
    \begin{equation}\label{eqn:num_assump_complexity_final_1}
        \norm{L_j^{-1} R_j - \mathcal{T} e^{\int_{jh}^{(j+1)h} A(s) ds}  } \leq \min\left\{ \frac{1}{2} \eta h e^{-\eta h}, \frac{\norm{u(T)}}{ \norm{u_0} + \max\norm{b(t)}/\eta} \frac{\eta h \epsilon}{128 }\right\}, 
    \end{equation}
    and 
    \begin{equation}\label{eqn:num_assump_complexity_final_2}
        \norm{ L_j^{-1} v_j - \int_{jh}^{(j+1)h} \mathcal{T} e^{\int_{s}^{(j+1)h} A(\tau) d\tau} b(s) ds  } \leq \frac{\norm{u(T)} \eta h \epsilon }{32} . 
    \end{equation}
    Then, we can obtain an $\epsilon$-approximation of the final state $u(T)/\norm{u(T)}$, using 
    \begin{equation}
        \widetilde{\Or}\left( \left( 1 + \max_j \norm{L_j} + \max_j \norm{R_j} \right) \left( 1+\max_j \norm{L_j^{-1}} \right)  \frac{\max_t \norm{u(t)}}{\norm{u(T)}} \sqrt{\frac{M+M_p}{M_p}}  \left(\frac{M}{\eta T} + M_p\right) \log\left( \frac{ 1 } {\epsilon} \right) \right)
    \end{equation}
    queries to the block-encoding of $\mathbf{A}_{M,M_p-1}$ and the state preparation oracle of $\ket{\mathbf{b}_{M,M_p-1}}$. 
\end{thm}
\begin{proof}
    Let $M = T/h$ and 
    \begin{equation}\label{eqn:proof_final_choice_epsilon}
        \epsilon_1 =  \frac{\norm{u(T)}}{ \norm{u_0} + \max\norm{b(t)}/\eta} \frac{\eta h \epsilon}{128 }  ,   \quad \epsilon_2 = \frac{\norm{u(T)} \eta h \epsilon }{32}. 
    \end{equation}
    Then the assumptions on the local operators become 
    \begin{equation}
        \norm{L_j^{-1} R_j - \mathcal{T} e^{\int_{jh}^{(j+1)h} A(s) ds}  } \leq \epsilon_1 , 
    \end{equation}
    and 
    \begin{equation}
        \norm{ L_j^{-1} v_j - \int_{jh}^{(j+1)h} \mathcal{T} e^{\int_{s}^{(j+1)h} A(\tau) d\tau} b(s) ds  } \leq \epsilon_2. 
    \end{equation}
    In the proof of~\cref{thm:complexity_single_step_inhomo}, we show that the time discretization error at step $k$ can be bounded as 
    \begin{equation}
        \norm{u_k - u(kh)} \leq k \epsilon_1 e^{-\eta (k-1)h/2} \norm{u_0} + \frac{\epsilon_1 h \max\norm{b(t)}}{(1-e^{-\eta h/2})^2} + \frac{\epsilon_2}{1-e^{-\eta h/2}}. 
    \end{equation}
    Using the facts that $1-e^{-x} \geq x/2$ for $x \in [0,1]$ and the function $x e^{-\eta (x-1)h/2}$ attains its maximum at $x = 2/(\eta h)$, we can further bound the time discretization error as 
    \begin{align}
        \norm{u_k - u(kh)} &\leq \frac{2 \norm{u_0} }{\sqrt{e}} \frac{\epsilon_1}{\eta h} + \frac{16\epsilon_1 \max\norm{b(t)}}{ \eta^2 h } + \frac{4 \epsilon_2}{ \eta h } \\
        & \leq \frac{16 \epsilon_1}{\eta h}  \left( \norm{u_0} + \frac{\max\norm{b(t)}}{\eta} \right) + \frac{4 \epsilon_2}{ \eta h }. 
    \end{align}
    We can verify that the choices of $\epsilon_1$ and $\epsilon_2$ in~\cref{eqn:proof_final_choice_epsilon} ensures 
    \begin{equation}\label{eqn:proof_final_complexity_num_error}
        \norm{ u_k - u(kh) } \leq \frac{\norm{u(T)}\epsilon}{4}. 
    \end{equation}
    \cref{eqn:proof_final_complexity_num_error} also holds for $k = M$. 
    Then the error in the normalized solutions becomes 
    \begin{align}\label{eqn:proof_final_complexity_eq2}
        \norm{\frac{u_M}{\norm{u_M}} - \frac{u(T)}{\norm{u(T)}}} & \leq \frac{2}{\norm{u(T)}} \norm{u_M - u(T)} 
        \leq \frac{\epsilon}{2}. 
    \end{align}

    Now let us determine the required accuracy in solving $\mathbf{A}_{M,M_p-1} \mathbf{u} = \mathbf{b}_{M,M_p-1}$. 
    Let $\mathbf{u} = \mathbf{A}_{M,M_p-1}^{-1} \mathbf{b}_{M,M_p-1}$ and $\ket{\mathbf{u}'}$ be the state produced by a quantum linear system algorithm up to error $\epsilon'$, in the sense that 
    \begin{equation}
        \norm{ \ket{\mathbf{u}'} - \mathbf{u}/\norm{\mathbf{u}} } \leq \epsilon'. 
    \end{equation}
    Let $\ket{\mathbf{u}'} = [u_0';u_1';\cdots;u_{M+M_p-1}']$ be partitioned in the same way as $\mathbf{u}$. 
    Then for any $k \geq M$, we have 
    \begin{equation}\label{eqn:proof_final_complexity_eq5}
        \norm{ u_k' - u_M / \norm{\mathbf{u}} } \leq \epsilon'. 
    \end{equation}
    Upon a successful measurement, we would get the state $\ket{u_k'}$, which satisfies 
    \begin{equation}\label{eqn:proof_final_complexity_eq1}
        \norm{ \ket{u_k'} - \ket{u_M} } = \norm{ \frac{u_k'}{\norm{u_k'}} - \frac{u_M / \norm{\mathbf{u}}}{\norm{u_M / \norm{\mathbf{u}}}} } \leq \frac{2\norm{\mathbf{u}}}{\norm{u_M}} \norm{ u_k' - u_M / \norm{\mathbf{u}} } \leq \frac{2\norm{\mathbf{u}}}{\norm{u_M}} \epsilon'. 
    \end{equation}
    Using~\cref{eqn:proof_final_complexity_num_error} and triangle inequality, we have, for any $0 \leq k \leq M$, 
    \begin{equation}
        \norm{u(kh)} - \frac{\norm{u(T)} \epsilon }{4} \leq \norm{u_k} \leq \max_t \norm{u(t)} + \frac{\norm{u(T)} \epsilon }{4}, 
    \end{equation}
    so 
    \begin{equation}\label{eqn:proof_final_complexity_eq7}
        \norm{u_M} \geq \norm{u(T)} (1-\epsilon/4), 
    \end{equation}
    and 
    \begin{align}
        \norm{\mathbf{u}} & \leq \sqrt{M+M_p} \max_k \norm{u_k} \leq \sqrt{M+M_p} \left(\max_t \norm{u(t)} + \frac{\norm{u(T)} \epsilon}{4} \right) \nonumber \\
        & \leq \sqrt{M+M_p} \max_t \norm{u(t)}\left( 1+ \frac{\epsilon}{4} \right). \label{eqn:proof_final_complexity_eq6}
    \end{align}
    Then~\cref{eqn:proof_final_complexity_eq1} can be further bounded as 
    \begin{equation}
        \norm{ \ket{u_k'} - \ket{u_M} } \leq  \frac{2 \sqrt{M+M_p} \max_t \norm{u(t)} \left(1 + \epsilon / 4 \right) }{\norm{u(T)} (1-\epsilon/4) } \epsilon' \leq \frac{4\sqrt{M+M_p}  \max_t \norm{u(t)}}{\norm{u(T)}} \epsilon'. 
    \end{equation}
    In order to bound this by $\epsilon/2$, it suffices to choose 
    \begin{equation}\label{eqn:proof_final_complexity_eq4}
        \epsilon' = \frac{\norm{u(T)}}{8\sqrt{M+M_p} \max_t \norm{u(t)} } \epsilon. 
    \end{equation}
    Then 
    \begin{equation}\label{eqn:proof_final_complexity_eq3}
        \norm{ \ket{u_k'} - \ket{u_M} } \leq \frac{\epsilon}{2}. 
    \end{equation}
    
    We have shown in~\cref{eqn:proof_final_complexity_eq2,eqn:proof_final_complexity_eq3} that, by solving the linear system up to $\epsilon'$ defined in~\cref{eqn:proof_final_complexity_eq4} and successfully measuring the index register, the produced quantum state is an $\epsilon$-approximation of the exact final state $\ket{u(T)}$. 
    According to~\cite{CostaAnYuvalEtAl2022}, solving the linear system requires $\Or( \kappa \log(1/\epsilon') )$ queries to the block-encoding of $\mathbf{A}_{M,M_p-1}$ and the state preparation oracle of $\mathbf{b}_{M,M_p-1}$, where $\kappa = \norm{\mathbf{A}_{M,M_p-1}} \norm{\mathbf{A}_{M,M_p-1}^{-1}}$ is the condition number of $\mathbf{A}_{M,M_p-1}$. 
    According to~\cref{lemma:condition-number}, we have 
    \begin{equation}
        \kappa = \Or\left( \left( 1 + \max_j \norm{L_j} + \max_j \norm{R_j} \right) \left( 1+\max_j \norm{L_j^{-1}} \right)  \left(\frac{M}{\eta T} + M_p\right) \right). 
    \end{equation}
    So the query complexity in each run of the algorithm before measurement is 
    \begin{equation}
        \Or\left(  \left( 1 + \max_j \norm{L_j} + \max_j \norm{R_j} \right) \left( 1+\max_j \norm{L_j^{-1}} \right)  \left(\frac{M}{\eta T} + M_p\right) \log\left( \frac{(M+M_p) \max_t \norm{u(t)} } {\norm{u(T)}\epsilon} \right) \right). 
    \end{equation}

    Now we estimate the number of repeats of solving the linear system. 
    The successful case is when the measurement outcome $k \geq M$. 
    To boost the successful probability to $\Omega(1)$ with amplitude amplification, we need $\Or(1/\sqrt{\sum_{k=M}^{M+M_p-1} \norm{u_k'}^2 })$ rounds. 
    For each $k \geq M$, according to~\cref{eqn:proof_final_complexity_eq5,eqn:proof_final_complexity_eq6,eqn:proof_final_complexity_eq7} and the definition of $\epsilon'$ in~\cref{eqn:proof_final_complexity_eq4}, we can bound 
    \begin{align}
        \norm{u_k'} & \geq \frac{\norm{u_M}}{\norm{\mathbf{u}}} - \epsilon' \geq \frac{\norm{u(T)} (1-\epsilon/4) }{\sqrt{M+M_p} \max_t \norm{u(t)}\left( 1+ \epsilon/4 \right)} -  \frac{\norm{u(T)}}{8\sqrt{M+M_p} \max_t \norm{u(t)} } \epsilon \nonumber \\
        & \geq \frac{\norm{u(T)}}{4\sqrt{M+M_p} \max_t \norm{u(t)} } . 
    \end{align}
    So the number of amplitude amplification rounds becomes 
    \begin{equation}
        \Or\left( \frac{1}{\sqrt{\sum_{k=M}^{M+M_p-1} \norm{u_k'}^2 } } \right) = \Or\left( \frac{\max_t \norm{u(t)}}{\norm{u(T)}} \sqrt{\frac{M+M_p}{M_p}} \right), 
    \end{equation}
    which contributes to another multiplicative factor in the overall complexity and completes the proof. 
\end{proof}

In~\cref{thm:complexity_final_state}, we can see two opposite effects of the parameter $M_p$. 
The factor $\sqrt{\frac{M+M_p}{M_p}}$ is due to the amplitude amplification, and will decrease as we choose larger $M_p$ to have more copies of the final solution. 
However, larger $M_p$ will increase the condition number of the linear system and thus the complexity of solving it, as shown in the factor $\frac{M}{\eta T} + M_p$. 
The following result shows the optimal choice of $M_p$ to best balance these two effects for large $T$. 

\begin{cor}\label{cor:complexity_final_state_optimal_Mp}
    Under the same conditions as~\cref{thm:complexity_final_state} and by choosing 
    \begin{equation}
        M_p = \lceil \frac{M}{\eta T} \rceil, 
    \end{equation}
    we can obtain an $\epsilon$-approximation of the final state $u(T)/\norm{u(T)}$, using 
    \begin{equation}
        \widetilde{\Or}\left( \left( 1 + \max_j \norm{L_j} + \max_j \norm{R_j} \right) \left( 1+\max_j \norm{L_j^{-1}} \right)  \frac{\max_t \norm{u(t)}}{\norm{u(T)}} \frac{M \sqrt{1+\eta T} }{\eta T} \log\left( \frac{ 1 } {\epsilon} \right) \right) 
    \end{equation}
    queries to the block-encoding of $\mathbf{A}_{M,M_p-1}$ and the state preparation oracle of $\ket{\mathbf{b}_{M,M_p-1}}$. 
\end{cor}
\begin{proof}
    It is a straightforward consequence of~\cref{thm:complexity_final_state} by noticing that $f(x) = \sqrt{\frac{M+x}{x}}  \left(\frac{M}{\eta T} + x\right)$ achieves its minimum at $x = \frac{2M}{\eta T (1 + \sqrt{1+8/(\eta T)}) }$.
\end{proof}

\section{Specific numerical methods}\label{sec:specific_methods}

We have established a complexity estimate for preparing a history state and a final state of a dissipative ODE in~\cref{sec:complexity}. 
Notice that the results in~\cref{sec:complexity} work for any single-step numerical method, yet the overall complexity in terms of $T$ and $\epsilon$ still remains unclear as the number of time steps $M$ still has implicit time and precision dependence. 
Here we discuss several examples of numerical methods and present their complexity estimates in a more explicit way. 
Examples include the state-of-the-art truncated Dyson series method, and two lower-order methods.  

Throughout this section, we assume the state preparation oracle $O_{u_0}$ for $\ket{u_0}$ such that 
\begin{equation}\label{eqn:def_oracle_u0}
    O_{u_0} \ket{0} = \ket{u_0}, 
\end{equation}
the input model $O_A$ for $A(t)$ such that
\begin{equation}
\label{eqn:def_oracle_A}
    O_A \ket{0} \ket{t} \ket{\psi} = \frac{1}{\alpha_A} \ket{0} \ket{t} A(t) \ket{\psi} + \ket{\perp},    
\end{equation}
and the input model $O_b$ for $b(t)$ such that 
\begin{equation}\label{eqn:def_oracle_b}
    O_b \ket{0} \ket{t} \ket{0} = \frac{1}{\alpha_b} \ket{0} \ket{t} \ket{b(t)} + \ket{\perp}. 
\end{equation}
Here $O_A$ is a time-dependent block-encoding of $A(t)$, and $O_b$ is a time-dependent state preparation oracle of $b(t)$. 
The normalization factors $\alpha_A$ and $\alpha_b$ satisfy $\alpha_A \geq \max_t \norm{A(t)}$ and $\alpha_b \geq \max_t \norm{b(t)}$. 
$\ket{t}$ represents an encoding of the time which usually refers to the index in specific methods.  
Such time-dependent version of the block-encoding model of $A(t)$ and state preparation of $b(t)$ have also been assumed in existing quantum algorithms for Hamiltonian simulation~\cite{LowWiebe2019} and general differential equations~\cite{FangLinTong2022,BerryCosta2022,AnChildsLin2023}.

\subsection{Truncated Dyson series method}\label{sec:Dyson}

The truncated Dyson series method is introduced in~\cite{BerryCosta2022} and achieves high-order convergence. 
In this method, we have 
\begin{equation}
    L_j = I, 
\end{equation}
\begin{equation}\label{eqn:Dyson_R}
    R_j = \sum_{k=0}^{K} \int_{jh}^{(j+1)h} d t_1  \int_{jh}^{t_1} d t_2 \cdots \int_{jh}^{t_{k-1}} d t_k A(t_1)A(t_2)\cdots A(t_k), 
\end{equation}
and 
\begin{equation}\label{eqn:Dyson_v}
    v_j = \sum_{k=1}^K \int_{jh}^{(j+1)h} d t_1  \int_{jh}^{t_1} d t_2 \cdots \int_{jh}^{t_{k-1}} d t_k A(t_1)A(t_2)\cdots A(t_{k-1})b(t_k).  
\end{equation}
Here $K$ is the truncation order in the Dyson series and only has a logarithmic dependence on the local truncation error, according to~\cite[Section 1]{BerryCosta2022}. 

Notice that in the actual implementation, integrals in~\cref{eqn:Dyson_R,eqn:Dyson_v} need to be further discretized with sufficiently many grid points, but this only introduces logarithmically many extra ancilla qubits and gates, and does not affect the overall number of queries to the input model of $A(t)$. 
For technical simplicity, here we will omit this further step of integral discretization as well as its discretization error, and refer to~\cite{BerryCosta2022} for more details.

We first discuss the complexity in the general inhomogeneous case, which is given in the following result. 

\begin{cor}\label{cor:Dyson_hist_inhomo}
    Consider using the truncated Dyson series method for solving the ODE~\cref{eqn:ODE} with $A(t)$ being a matrix such that $A(t) + A(t)^{\dagger} \leq -2\eta < 0$. 
    Let $\epsilon > 0$ be theon target error, $T \geq \eta^{-1}$ be the evolution time, $\alpha_A \geq \max\norm{A(t)}$ be the block-encoding factor of $A(t)$, and $\alpha_b \geq \max\norm{b(t)}$ be the normalization factor of $b(t)$. 
    Then, we can obtain an $\epsilon$-approximation of 
    \begin{enumerate}
        \item the history state $\mathbf{u}_{\text{exact}}/\norm{\mathbf{u}_{\text{exact}}}$ using 
    \begin{equation}
        \Or\left( \frac{\alpha_A}{\eta } \log\left( \frac{ (\alpha_A + \alpha_b) T}{ \norm{u_0} \eta \epsilon }  \right)  \log\left(\frac{1}{\epsilon}\right) \right)
    \end{equation}
    queries to the time-dependent block-encoding $O_A$ of $A(t)$ and 
    \begin{equation}
         \Or\left( \frac{\alpha_A}{\eta } \log\left(\frac{1}{\epsilon}\right) \right)
    \end{equation}
    queries to the state preparation $O_{u_0}$ of $\ket{u_0}$ and $O_b$ of $\ket{b(t)}$. 
        \item the final state $u(T)/\norm{u(T)}$ using 
    \begin{equation}
        \widetilde{\Or}\left( \frac{\max_t \norm{u(t)}}{\norm{u(T)}} \frac{ \alpha_A \sqrt{T} }{\sqrt{\eta}} \log\left( \frac{\norm{u_0}+\alpha_b}{\norm{u(T)}} \frac{\alpha_A}{\eta \epsilon} \right) \log\left( \frac{ 1 } {\epsilon} \right) \right)
    \end{equation}
    queries to the time-dependent block-encoding $O_A$ of $A(t)$ and 
    \begin{equation}
         \widetilde{\Or}\left( \frac{\max_t \norm{u(t)}}{\norm{u(T)}} \frac{\alpha_A \sqrt{T} }{\sqrt{\eta}} \log\left( \frac{ 1 } {\epsilon} \right) \right) 
    \end{equation}
    queries to the state preparation $O_{u_0}$ of $\ket{u_0}$ and $O_b$ of $\ket{b(t)}$. 
    \end{enumerate}
\end{cor}

\begin{proof}
    We first consider the history state preparation. 
    According to~\cite[Section 3]{BerryCosta2022}, constructing a block-encoding of $\mathbf{A}_{M,M_p-1}$ needs $K$ calls to $O_A$, and constructing a state preparation oracle of $\ket{\mathbf{b}_{M,M_p-1}}$ needs $K-1$ calls to $O_A$, one call to $O_b$, and one call to $O_{u_0}$. 
    Therefore,~\cref{thm:complexity_single_step_inhomo} tells that the overall query complexity is 
    \begin{equation}\label{eqn:proof_Dyson_hist_inhomo_eq3}
        \Or\left( \left( 1 + \max_j \norm{R_j} \right) \frac{KM}{\eta T} \log\left(\frac{1}{\epsilon}\right) \right)
    \end{equation}
    queries to $O_A$, and 
    \begin{equation}\label{eqn:proof_Dyson_hist_inhomo_eq4}
        \Or\left( \left( 1 + \max_j \norm{R_j} \right) \frac{M}{\eta T} \log\left(\frac{1}{\epsilon}\right) \right)
    \end{equation}
    queries to $O_{u_0}$ and $O_b$. 

    We need to choose $K$ and $M$ such that the assumptions in~\cref{thm:complexity_single_step_inhomo} hold. 
    Notice that, under the assumption $\eta h \leq 1$ which will be satisfied by the choice of $h$ specified later, the first term on the right hand side of~\cref{eqn:num_assump_complexity_inhomo_1} is always larger than the second term. 
    So we need to choose $K$ and $M$ such that 
    \begin{equation}\label{eqn:proof_Dyson_hist_inhomo_eq1}
        \norm{ R_j - \mathcal{T} e^{\int_{jh}^{(j+1)h} A(s) ds}  } \leq \Or\left(  \frac{ \eta^{3/2} h \epsilon }{  \sqrt{1+\frac{T\max\norm{b}^2 }{\eta \norm{u_0}^2 }}  \sqrt{\max\norm{A(t)} + \frac{\max\norm{b(t)}}{\norm{u_0}} } }   \right), 
    \end{equation}
    and 
    \begin{equation}\label{eqn:proof_Dyson_hist_inhomo_eq2}
        \norm{ v_j - \int_{jh}^{(j+1)h} \mathcal{T} e^{\int_{s}^{(j+1)h} A(\tau) d\tau} b(s) ds  } \leq \Or\left( \frac{\norm{u_0} \eta h \epsilon}{ \sqrt{T} \sqrt{ \max\norm{A(t)} + \frac{\max\norm{b(t)}}{\norm{u_0}} } } \right). 
    \end{equation}

    We first choose 
    \begin{equation}
        M = 2 \alpha_A T, 
    \end{equation}
    then $\alpha_A h = 1/2$ and thus the block-encoding of the Dyson series is well-behaved~\cite{BerryCosta2022}. 
    As a result, by the definition of $R_j$ in~\cref{eqn:Dyson_R}, we have 
    \begin{equation}
        \norm{R_j} \leq \sum_{k=0}^K \frac{h^k}{k!} \alpha_A^k \leq e^{\alpha_A h} = \Or(1). 
    \end{equation}
    Furthermore, according to~\cite[Eq.(6)]{BerryCosta2022}, the local truncation error can be bounded as 
    \begin{equation}\label{eqn:proof_Dyson_hist_inhomo_eq5}
         \norm{ R_j - \mathcal{T} e^{\int_{jh}^{(j+1)h} A(s) ds}  } \leq \Or\left( \frac{ (\alpha_A h)^{K+1} }{(K+1)!} \right) \leq \Or\left( \frac{ 1 }{(K+1)!} \right) \leq  \Or\left( e^{-K} \right). 
    \end{equation}
    In order to bound this error according to~\cref{eqn:proof_Dyson_hist_inhomo_eq1}, it suffices to choose 
    \begin{equation}
        K = \Or\left(  \log\left( \frac{  \sqrt{1+\frac{T\max\norm{b}^2 }{\eta \norm{u_0}^2 }}  \sqrt{\max\norm{A(t)} + \frac{\max\norm{b(t)}}{\norm{u_0}} } }{ \eta^{3/2} h \epsilon }  \right)  \right) = \Or\left(  \log\left( \frac{ (\alpha_A + \alpha_b) T}{ \norm{u_0} \eta \epsilon }  \right) \right). 
    \end{equation}
    For the inhomogeneous error,~\cite[Eq.(14)]{BerryCosta2022} suggests that 
    \begin{equation}\label{eqn:proof_Dyson_hist_inhomo_eq6}
        \norm{  v_j - \int_{jh}^{(j+1)h} \mathcal{T} e^{\int_{s}^{(j+1)h} A(\tau) d\tau} b(s) ds  } \leq \Or\left( \frac{\alpha_A^K h^{K+1} \alpha_b}{ (K+1)! } \right) \leq \Or\left( \frac{\alpha_b h }{ (K+1)! } \right) \leq \Or\left( \alpha_b h e^{-K} \right). 
    \end{equation}
    In order to bound this error according to~\cref{eqn:proof_Dyson_hist_inhomo_eq2}, it suffices to choose 
    \begin{equation}
        K = \Or\left( \log\left( \frac{ \alpha_b \sqrt{T} \sqrt{ \max\norm{A(t)} + \frac{\max\norm{b(t)}}{\norm{u_0}} } }{\norm{u_0} \eta  \epsilon} \right) \right) = \Or\left(  \log\left( \frac{ (\alpha_A + \alpha_b) T}{ \norm{u_0} \eta \epsilon }  \right) \right),  
    \end{equation}
    which is asymptotically the same as the choice of $K$ for bounding the homogeneous error. 
    Plugging the choices of $M$ and $K$ back to~\cref{eqn:proof_Dyson_hist_inhomo_eq3,eqn:proof_Dyson_hist_inhomo_eq4} yields the claimed complexity estimates for history state preparation. 

    We now move on to the final state preparation, where the proof is quite similar. 
    \cref{cor:complexity_final_state_optimal_Mp} tells that the overall query complexity is 
    \begin{align}
        & \widetilde{\Or}\left( \left( 1 + \max_j \norm{R_j} \right)  \frac{\max_t \norm{u(t)}}{\norm{u(T)}} \frac{K M \sqrt{1+\eta T} }{\eta T} \log\left( \frac{ 1 } {\epsilon} \right) \right) \nonumber \\
        =~ & \widetilde{\Or}\left( \left( 1 + \max_j \norm{R_j} \right)  \frac{\max_t \norm{u(t)}}{\norm{u(T)}} \frac{K M }{\sqrt{\eta T}} \log\left( \frac{ 1 } {\epsilon} \right) \right) \label{eqn:proof_Dyson_hist_inhomo_eq7}
    \end{align}
    queries to $O_A$, and 
    \begin{align}
        &\widetilde{\Or}\left( \left( 1 + \max_j \norm{R_j} \right)  \frac{\max_t \norm{u(t)}}{\norm{u(T)}} \frac{M \sqrt{1+\eta T} }{\eta T} \log\left( \frac{ 1 } {\epsilon} \right) \right) \nonumber\\
        =~ &\widetilde{\Or}\left( \left( 1 + \max_j \norm{R_j} \right)  \frac{\max_t \norm{u(t)}}{\norm{u(T)}} \frac{M }{\sqrt{\eta T}} \log\left( \frac{ 1 } {\epsilon} \right) \right) \label{eqn:proof_Dyson_hist_inhomo_eq8}
    \end{align}
    queries to $O_{u_0}$ and $O_b$. 
    So we need to determine the choice of $K$ and $M$ in this case. 
    We can use the same choice $M = 2\alpha_A T$. 
    Then $\norm{R_j}$'s are still $\Or(1)$, and we still have the local truncation error bound in~\cref{eqn:proof_Dyson_hist_inhomo_eq5,eqn:proof_Dyson_hist_inhomo_eq6}. 
    To satisfy the assumptions in~\cref{thm:complexity_final_state} and~\cref{cor:complexity_final_state_optimal_Mp}, it suffices to require 
    \begin{equation}
        \Or(e^{-K}) \leq \Or\left( \frac{\norm{u(T)}\eta h \epsilon}{ \norm{u_0} + \max\norm{b(t)}/\eta}  \right), 
    \end{equation}
    and 
    \begin{equation}
        \Or(\alpha_b h e^{-K}) \leq \Or\left( \norm{u(T)} \eta h \epsilon  \right). 
    \end{equation}
    By solving these two inequalities, we can derive a sufficient condition for $K$ to be 
    \begin{equation}
        K = \Or\left(  \log\left( \frac{\norm{u_0}+\alpha_b}{\norm{u(T)}} \frac{\alpha_A}{\eta \epsilon} \right)  \right). 
    \end{equation}
    Plugging these choices back to~\cref{eqn:proof_Dyson_hist_inhomo_eq7,eqn:proof_Dyson_hist_inhomo_eq8} completes the second part of the proof. 
\end{proof}

In the special homogeneous case, the complexity for preparing a history state can be slightly improved by applying the tighter estimate in~\cref{thm:complexity_single_step_homo}. 
We state its complexity in the following result and give its proof in~\cref{app:proof_Dyson_hist_homo}.

\begin{cor}\label{cor:Dyson_hist_homo}
    Consider using the truncated Dyson series method for solving the ODE~\cref{eqn:ODE} with $b(t) \equiv 0$ and $A(t)$ being a matrix such that $A(t) + A(t)^{\dagger} \leq -2\eta < 0$. 
    Let $\epsilon > 0$ be the target error, $T \geq (\max\norm{A(t)})^{-1}$ be the evolution time, and $\alpha_A \geq \max\norm{A(t)}$ be the block-encoding factor of $A(t)$. 
    Then, we can obtain an $\epsilon$-approximation of the history state $\mathbf{u}_{\text{exact}}/\norm{\mathbf{u}_{\text{exact}}}$, using 
    \begin{equation}
        \Or\left(  \frac{\alpha_A}{\eta } \log\left( \frac{ \alpha_A }{\eta \epsilon } \right) \log\left( \frac{1}{\epsilon} \right)  \right)
    \end{equation}
    queries to the time-dependent block-encoding $O_A$ of $A(t)$ and 
    \begin{equation}
        \Or\left( \frac{\alpha_A}{\eta} \log\left( \frac{1}{\epsilon} \right)  \right)
    \end{equation}
    queries to the state preparation $O_{u_0}$ of $\ket{u_0}$. 
\end{cor}

\subsection{Lower-order methods}\label{sec:lower_order}

We have shown that the truncated Dyson series method can scale $\Or(\text{poly}\log(T/\epsilon))$ for history state preparation and $\widetilde{\Or}(\sqrt{T}\text{poly}\log(1/\epsilon))$ for final state preparation. 
The dependence on time $T$ achieves fast-forwarding compared to the previous linear scaling. 
In this subsection, we will show that such a fast-forwarded time dependence can also be achieved by much simpler lower-order methods such as forward Euler method and trapezoidal rule.

To highlight our main point and simplify the technical analysis, throughout this subsection, we will only focus on the explicit complexity in terms of time $T$ and error $\epsilon$, and treat all the other parameters as constants and absorb them into the big-O notation. 
We will require the time-discretization $h = T/N$ to satisfy $h \alpha_A = \mathcal{O}(1)$. 
Thus we are able to 
construct the block-encoding of $\mathbf{A}_{M,M_p-1}$ defined in~\cref{eqn:def_A_padding} corresponding
to different numerical schemes with a $\mathcal{O}(1)$ normalization factor.
The constructing process is postponed to~\cref{appendix:block-encodings}.

\subsubsection{Forward Euler method}
The forward Euler method is to approximate $\mathcal{T}e^{\int_{jh}^{(j+1)h}A(s) \mathrm{d}s}$ by
$I + h A(jh)$. According to the analysis in ~\cref{appendix:ea}, we know
\begin{equation}
    \left\|\mathcal{T}e^{\int_{jh}^{(j+1)h} A(s) \mathrm{d}s} -  \left(I + h A(jh)\right)\right\| = 
    \mathcal{O}(h^2)
\end{equation}
and
\begin{equation}
    \left\|hb((j+1)h) - \int_0^h\mathcal{T}e^{\int_{s}^{(j+1)h} A(t) \mathrm{d}t}b(s)\mathrm{d}t\right\| = 
    \mathcal{O}(h^2).
\end{equation}
We first present its complexity estimate for the general inhomogeneous ODE. 

\begin{cor}\label{cor:Euler_hist}
    To solve the ODE~\cref{eqn:ODE} with $A(t)$ satisfying the stability condition~\cref{eqn:stability} by the forward Euler method, 
    let $\epsilon$ be the desired precision, $T > \left(\max \|A(t)\|\right)^{-1}$
    be the evolution time, 
    then we can obtain an $\epsilon$-approximation of
    \begin{enumerate}
        \item the history state $\mathbf{u}_{\text{exact}}/\|\mathbf{u}_{\text{exact}}\|$ using
        \begin{equation}
            \mathcal{O}\left( \frac{\sqrt{T}}{\epsilon}\log\left(\frac{1}{\epsilon}\right)\right)
        \end{equation}
        queries to the time-dependent block-encoding $O_A$ of $A(t)$ and the state preparation oracle.
        \item the final state $u(T)/\|u(T)\|$ using
        \begin{equation}
            \widetilde{\mathcal{O}}\left( \frac{\sqrt{T}}{\epsilon}\log\left(\frac{1}{\epsilon}\right)\right)
        \end{equation}
        queries to the time-dependent block-encoding $O_A$ of $A(t)$ and the state preparation oracle.
    \end{enumerate}
\end{cor}

\begin{proof}
    According to the discussion in~\cref{appendix:block-encodings}, we need one query of $O_A$ to
    construct the block-encoding of $\mathbf{A}_{M,M_p-1}$. 
    \cref{thm:complexity_single_step_inhomo} tells us that we need ~\cref{eqn:num_assump_complexity_inhomo_1} and~\cref{eqn:num_assump_complexity_inhomo_2} hold for the numerical scheme.

    In the inhomogeneous case,
    the first requirement is
    \begin{align}
    & \norm{L(j,h)^{-1} R(j,h) - \mathcal{T} e^{\int_{jh}^{(j+1)h} A(s) ds}  } \nonumber \\
    \leq~ & \min\left\{ \frac{1}{2} \eta h e^{-\eta h}, \frac{ \eta^{3/2} h \epsilon }{ 144\sqrt{2} \sqrt{1+\frac{T\max\norm{b(t)}^2 }{\eta\|u_0\|^2}}  \sqrt{\max\norm{A(t)} + \frac{\max\norm{b(t)}}{\|u_0\|} } } \right\}, 
    \end{align}
    and what we have is
    \begin{equation}
        \norm{L(j,h)^{-1} R(j,h) - \mathcal{T} e^{\int_{jh}^{(j+1)h} A(s) ds}  } = \mathcal{O}(h^2),
    \end{equation}
    so we can just require
    \begin{equation}
    \label{eqn:fe-req1}
        h \lesssim \min\left\{\eta,
        \frac{ \eta^{3/2} \epsilon }{\sqrt{1+\frac{T\max\norm{b(t)}^2 }{\eta\norm{u_0}^2}}  \sqrt{\alpha_A + \frac{\max\norm{b(t)}}{\|u_0\|} } }
        \right\} = \Or\left( \frac{\epsilon}{\sqrt{T}} \right).
    \end{equation}

    For the second requirement
    \begin{equation}
        \norm{ L(j,h)^{-1} v(j,h) - \int_{jh}^{(j+1)h} \mathcal{T} e^{\int_{s}^{(j+1)h} A(\tau) d\tau} b(s) ds  } \leq \frac{\|u_0\|\eta h \epsilon}{ 72\sqrt{2} \sqrt{T} \sqrt{ \max_t\|A(t)\| + \frac{\max\norm{b(t)}}{\|u_0\|}  } },
    \end{equation}
    we would require
    \begin{equation}
    \label{eqn:fe-req2}
        h \lesssim \frac{\|u_0\|\eta \epsilon}{\sqrt{T} \sqrt{\alpha_A + \frac{\max\norm{b(t)}}{\|u_0\|}  }} = \Or\left( \frac{\epsilon}{\sqrt{T}} \right). 
    \end{equation}
    For the complexity of $T$ and $\epsilon$, equation~\cref{eqn:fe-req1} and equation~\cref{eqn:fe-req2} indicate that
    \begin{equation}
        M = \mathcal{O}\left(T^{3/2}\frac{1}{\epsilon}\right).
    \end{equation}
    So according to~\cref{thm:complexity_single_step_inhomo}, we will need
    \begin{equation}
    \mathcal{O}\left(T^{1/2}\frac{1}{\epsilon}\log\left(\frac{1}{\epsilon}\right)\right)
    \end{equation}
    queries to $O_A$ and the state preparation oracle that produces $\ket{\mathbf{b}}$.

    For the final state preparation, the first requirement in~\cref{thm:complexity_final_state} is
    \begin{equation}
        \norm{L_j^{-1} R_j - \mathcal{T} e^{\int_{jh}^{(j+1)h} A(s) ds}  } \leq \min\left\{ \frac{1}{2} \eta h e^{-\eta h}, \frac{\norm{u(T)}}{ \norm{u_0} + \max\norm{b(t)}/\eta} \frac{\eta h \epsilon}{128 }\right\}, 
    \end{equation}
    and what we have is
    \begin{equation}
        \norm{L(j,h)^{-1} R(j,h) - \mathcal{T} e^{\int_{jh}^{(j+1)h} A(s) ds}  } = \mathcal{O}(h^2).
    \end{equation}
    So we can just require
    \begin{equation}
        h \lesssim \min\left\{\eta,
        \frac{\norm{u(T)}\eta \epsilon}{\norm{u_0} + \max\norm{b(t)}/\eta} 
        \right\} = \Or\left( \epsilon \right). 
    \end{equation}

    For the second requirement
    \begin{equation}
        \norm{ L_j^{-1} v_j - \int_{jh}^{(j+1)h} \mathcal{T} e^{\int_{s}^{(j+1)h} A(\tau) d\tau} b(s) ds  } \leq \frac{\norm{u(T)} \eta h \epsilon }{32},
    \end{equation}
    we would require
    \begin{equation}
        h \lesssim \|u(T)\|\eta\epsilon = \Or(\epsilon) .
    \end{equation}
    It is clear that
    \begin{equation}
        M = \mathcal{O}\left(\frac{T}{\epsilon}\right). 
    \end{equation}
    So according to~\cref{cor:complexity_final_state_optimal_Mp}, we will need
    \begin{equation}
        \widetilde{\mathcal{O}}\left(T^{1/2}\frac{1}{\epsilon}\log\left(\frac{1}{\epsilon}\right)\right)
    \end{equation}
    queries to $O_A$ and the state preparation oracle.
\end{proof}

\cref{cor:Euler_hist} shows that complexities of the forward Euler method for history state and final state are the same $\Or(\sqrt{T})$. 
For homogeneous ODE, the complexity for history state preparation can be further improved, as shown in the next result. 

\begin{cor}\label{cor:Euler_homo}
    To solve the ODE~\cref{eqn:ODE} with $A(t)$ satisfying the stability condition~\cref{eqn:stability} and $b(t) \equiv 0$ by the forward Euler method, 
    let $\epsilon$ be the desired precision, $T > \left(\max \|A(t)\|\right)^{-1}$
    be the evolution time, 
    then we can obtain an $\epsilon$-approximation of the history state $\mathbf{u}_{\text{exact}}/\|\mathbf{u}_{\text{exact}}\|$
    using
    \begin{equation}
        \mathcal{O}\left(\frac{1}{\epsilon}\log\left(\frac{1}{\epsilon}\right)\right)
    \end{equation}
    queries to $O_A$ and the state preparation oracle.
    \end{cor}
\begin{proof}
    According to~\cref{thm:complexity_single_step_homo}, for the homogeneous case, we only require that
    \begin{equation}
        \norm{L_j^{-1} R_j - \mathcal{T} e^{\int_{jh}^{(j+1)h} A(s) ds}  } \leq \min \left\{ \frac{1}{2}\eta h e^{-\eta h}, \frac{ \eta^{3/2} h \epsilon }{ 32 \sqrt{  \max \norm{A(t)}} } \right\}.  
    \end{equation}
    This leads to 
    \begin{equation}
        h \lesssim \min\left\{\eta, \frac{\eta^{3/2}\epsilon}{\alpha_A^{1/2}}\right\} = \Or(\epsilon), 
    \end{equation}
    meaning
    \begin{equation}
        M = \mathcal{O}\left( \frac{T}{\epsilon}\right).
    \end{equation}
    So according to~\cref{thm:complexity_single_step_homo}, there is no dependence on $T$, and the dependence on $\epsilon$ is 
    $\mathcal{O}\left(\frac{1}{\epsilon}\log\left(\frac{1}{\epsilon}\right)\right)$.
    
\end{proof}

\subsubsection{Trapezoidal rule}

The trapezoidal rule is to approximate $\mathcal{T}e^{\int_{jh}^{(j+1)h}A(s) \mathrm{d}s}$ by 
$\left(I - \frac{h}{2} A((j+1)h)\right)^{-1}\left(I + \frac{h}{2}A(jh)\right)$. According to the analysis in~\cref{appendix:ea}, we know
\begin{equation}
    \left\|\mathcal{T}e^{\int_{jh}^{(j+1)h} A(s) \mathrm{d}s} -  \left(I - \frac{h}{2} A((j+1)h)\right)^{-1}\left(I + \frac{h}{2}A(jh)\right)\right\| = 
    \mathcal{O}(h^3)
\end{equation}
and
\begin{equation}
    \left\|\left(I - \frac{h}{2} A((j+1)h)\right)^{-1} \frac{h}{2}(b(h) + b(0)) - \int_0^h\mathcal{T}e^{\int_{s}^{(j+1)h} A(t) \mathrm{d}t}b(s)\mathrm{d}t\right\| = 
    \mathcal{O}(h^3).
\end{equation}
\begin{cor}\label{cor:trap_hist}
    To solve the ODE~\cref{eqn:ODE} with $A(t)$ satisfying the stability condition~\cref{eqn:stability} by the trapezoidal rule, 
    let $\epsilon$ be the desired precision, $T > \left(\max \|A(t)\|\right)^{-1}$
    be the evolution time, 
    then we can obtain an $\epsilon$-approximation of
    \begin{enumerate}
        \item the history state $\mathbf{u}_{\text{exact}}/\|\mathbf{u}_{\text{exact}}\|$ using
        \begin{equation}
            \mathcal{O}\left(T^{1/4} \sqrt{\frac{1}{\epsilon}}\log\left(\frac{1}{\epsilon}\right)\right)
        \end{equation}
        queries to the time-dependent block-encoding $O_A$ of $A(t)$ and the state preparation oracle.
        \item the final state $u(T)/\|u(T)\|$ using
        \begin{equation}
            \widetilde{\mathcal{O}}\left(T^{1/2} \sqrt{\frac{1}{\epsilon}}\log\left(\frac{1}{\epsilon}\right)\right)
        \end{equation}
        queries to the time-dependent block-encoding $O_A$ of $A(t)$ and the state preparation oracle.
    \end{enumerate}
\end{cor}

\begin{proof}
    According to the discussion in~\cref{appendix:block-encodings}, we need two queries of $O_A$ to
    construct the block-encoding of $\mathbf{A}_{M,M_p-1}$. 
    \cref{thm:complexity_single_step_inhomo} tells us that we need ~\cref{eqn:num_assump_complexity_inhomo_1} and~\cref{eqn:num_assump_complexity_inhomo_2} hold for the numerical scheme.
    In the inhomogeneous case,
    the first requirement is
    \begin{align}
    & \norm{L(j,h)^{-1} R(j,h) - \mathcal{T} e^{\int_{jh}^{(j+1)h} A(s) ds}  } \nonumber \\
    \leq~ & \min\left\{ \frac{1}{2} \eta h e^{-\eta h}, \frac{ \eta^{3/2} h \epsilon }{ 144\sqrt{2} \sqrt{1+\frac{T\max\norm{b(t)}^2 }{\eta\|u_0\|^2}}  \sqrt{\max\norm{A(t)} + \frac{\max\norm{b(t)}}{\|u_0\|} } } \right\}, 
    \end{align}
    and what we have is
    \begin{equation}
        \norm{L(j,h)^{-1} R(j,h) - \mathcal{T} e^{\int_{jh}^{(j+1)h} A(s) ds}  } = \mathcal{O}(h^3),
    \end{equation}
    so we can just require
    \begin{equation}
    \label{eqn:tr-req1}
        h \lesssim \min\left\{\sqrt{\eta},
        \left(\frac{ \eta^{3/2} \epsilon }{\sqrt{1+\frac{T\max\norm{b(t)}^2 }{\eta\norm{u_0}^2}}  \sqrt{\alpha_A + \frac{\max\norm{b(t)}}{\|u_0\|} } }\right)^{1/2}
        \right\} = \Or\left( \frac{\epsilon^{1/2}}{T^{1/4}} \right).
    \end{equation}

    For the second requirement
    \begin{equation}
        \norm{ L(j,h)^{-1} v(j,h) - \int_{jh}^{(j+1)h} \mathcal{T} e^{\int_{s}^{(j+1)h} A(\tau) d\tau} b(s) ds  } \leq \frac{\|u_0\|\eta h \epsilon}{ 72\sqrt{2} \sqrt{T} \sqrt{ \max_t\|A(t)\| + \frac{\max\norm{b(t)}}{\|u_0\|}  } },
    \end{equation}
    we would require
    \begin{equation}
    \label{eqn:tr-req2}
        h \lesssim \left(\frac{\|u_0\|\eta \epsilon}{\sqrt{T} \sqrt{\alpha_A + \frac{\max\norm{b(t)}}{\|u_0\|}  }}\right)^{1/2} = \Or\left( \frac{\epsilon^{1/2}}{T^{1/4}} \right). 
    \end{equation}
    For the complexity of $T$ and $\epsilon$, equation~\cref{eqn:tr-req1} and equation~\cref{eqn:tr-req2} indicate that
    \begin{equation}
        M = \mathcal{O}\left(T^{5/4}\frac{1}{\sqrt{\epsilon}}\right).
    \end{equation}
    So according to theorem~\cref{thm:complexity_single_step_inhomo}, we will need
    \begin{equation}
    \mathcal{O}\left(T^{1/4}\frac{1}{\sqrt{\epsilon}}\log\left(\frac{1}{\epsilon}\right)\right)
    \end{equation}
    queries to $O_A$ and the state preparation oracle.
    
    For the final state preparation, the first requirement in~\cref{thm:complexity_final_state} is
    \begin{equation}
        \norm{L_j^{-1} R_j - \mathcal{T} e^{\int_{jh}^{(j+1)h} A(s) ds}  } \leq \min\left\{ \frac{1}{2} \eta h e^{-\eta h}, \frac{\norm{u(T)}}{ \norm{u_0} + \max\norm{b(t)}/\eta} \frac{\eta h \epsilon}{128 }\right\}, 
    \end{equation}
    and what we have is
    \begin{equation}
        \norm{L(j,h)^{-1} R(j,h) - \mathcal{T} e^{\int_{jh}^{(j+1)h} A(s) ds}  } = \mathcal{O}(h^3),
    \end{equation}
    so we can just require
    \begin{equation}
        h \lesssim \min\left\{\sqrt{\eta},
        \left(\frac{\norm{u(T)}\eta \epsilon}{(\norm{u_0} + \max\norm{b(t)}/\eta)} \right)^{1/2}
        \right\} = \Or(\sqrt{\epsilon}). 
    \end{equation}

    For the second requirement
    \begin{equation}
        \norm{ L_j^{-1} v_j - \int_{jh}^{(j+1)h} \mathcal{T} e^{\int_{s}^{(j+1)h} A(\tau) d\tau} b(s) ds  } \leq \frac{\norm{u(T)} \eta h \epsilon }{32},
    \end{equation}
    we would require
    \begin{equation}
        h \lesssim \left(\|u(T)\| \eta \epsilon\right)^{1/2} = \Or(\sqrt{\epsilon}). 
    \end{equation}
    It is clear then 
    \begin{equation}
        M = \mathcal{O}\left(T\frac{1}{\sqrt{\epsilon}}\right).
    \end{equation}
    So according to~\cref{cor:complexity_final_state_optimal_Mp}, we will need
    \begin{equation}
        \widetilde{\mathcal{O}}\left(T^{1/2}\frac{1}{\sqrt{\epsilon}}\log\left(\frac{1}{\epsilon}\right)\right)
    \end{equation}
    queries to $O_A$ and the state preparation oracle.
\end{proof}

In the homogeneous case, we have the following further improved complexity estimate for history state preparation. 

\begin{cor}\label{cor:trap_final}
    To solve the ODE~\cref{eqn:ODE} with $A(t)$ satisfying the stability condition~\cref{eqn:stability} and $b(t) \equiv 0$ by the trapezoidal rule, 
    let $\epsilon$ be the desired precision, $T > \left(\max \|A(t)\|\right)^{-1}$
    be the evolution time, then we can obtain an $\epsilon$-approximation of the history state $\mathbf{u}_{\text{exact}}/\|\mathbf{u}_{\text{exact}}\|$
    using 
    \begin{equation}
        \mathcal{O}\left(\frac{1}{\sqrt{\epsilon}}\log\left(\frac{1}{\epsilon}\right)\right)    
    \end{equation}
    queries to $O_A$ and the state preparation oracle. 
\end{cor}
\begin{proof}
    According to~\cref{thm:complexity_single_step_homo}, for the homogeneous case,
    we only require that
    \begin{equation}
        \norm{L_j^{-1} R_j - \mathcal{T} e^{\int_{jh}^{(j+1)h} A(s) ds}  } \leq \min \left\{ \frac{1}{2}\eta h e^{-\eta h}, \frac{ \eta^{3/2} h \epsilon }{ 32 \sqrt{  \max \norm{A(t)}} } \right\}, \quad \forall\; 0 \leq j \leq T/h-1.   
    \end{equation}
    This leads to 
    \begin{equation}
        h \lesssim \min\left\{\sqrt{\eta}, \sqrt{\frac{\eta^{3/2}\epsilon}{\alpha_A^{1/2}}}\right\} = \Or\left( \sqrt{\epsilon} \right), 
    \end{equation}
    meaning
    \begin{equation}
        M = \mathcal{O}\left(T \frac{1}{\sqrt{\epsilon}}\right).
    \end{equation}
    So according to~\cref{thm:complexity_single_step_homo}, there is no dependence on $T$, and the dependence on $\epsilon$ is 
    $\mathcal{O}\left(\frac{1}{\sqrt{\epsilon}}\log\left(\frac{1}{\epsilon}\right)\right)$. 
\end{proof}

\section{Applications}\label{sec:applications}

In this section, we discuss the complexity of quantum algorithm for some applications of dissipative ODEs, including quantum dynamics with non-Hermitian Hamiltonian and generalized heat equation. 
For the best query complexity, in this section we only focus on the complexity of using the truncated Dyson series method.

\subsection{Quantum dynamics with non-Hermitian Hamiltonian}

General form of quantum dynamics with non-Hermitian Hamiltonian is given as
\begin{equation}\label{eqn:nonH_dynamics}
    i \frac{du(t)}{dt} = ( H(t) + i L(t) ) u(t), \quad u(0) = \ket{u_0}. 
\end{equation}
Here the entire matrix $H(t) + i L(t)$ is called the non-Hermitian Hamiltonian, where both $H(t)$ and $L(t)$ are Hermitian. 
\cref{eqn:nonH_dynamics} is a generalization of the time-dependent Hamiltonian simulation problem. 
The $H(t)$ matrix is the standard (Hermitian) Hamiltonian in a closed quantum system, and $i L(t)$ represents the non-Hermitian correction term to the system. 

To apply our fast-forwarded algorithm, we assume~\cref{eqn:nonH_dynamics} to be dissipative, namely that there exists a constant $\eta >0$ such that $L(t) \leq -\eta < 0$. 
We suppose that we are given access to the time-dependent block-encodings $O_L$ and $O_H$ of $L(t)$ and $H(t)$, respectively. 
The block-encoding of $H(t) + i L(t)$ can be easily constructed by linearly adding them together using the linear combination of unitaries (LCU) technique~\cite{ChildsWiebe2012,ChildsKothariSomma2017}, and this only requires one query to each of the block-encodings. 

Notice that~\cref{eqn:nonH_dynamics} is a homogeneous ODE, so the dynamics decays exponentially and we only focus on the task of history state preparation. 
Quantum complexity is given in the following result, which is just a reformulation of~\cref{cor:Dyson_hist_homo} and has no explicit time dependence. 

\begin{cor}\label{cor:non_Hermitian}
    Consider using the truncated Dyson series method for simulating dissipative quantum dynamics with non-Hermitian Hamiltonian in~\cref{eqn:nonH_dynamics} with $L(t) \leq -\eta < 0$. 
    Let $\epsilon > 0$ be the target error, $T$ be the evolution time, and $\alpha_L \geq \max\norm{L(t)}$ and $\alpha_H \geq \max\norm{H(t)}$ be the block-encoding factors of $L(t)$ and $H(t)$, respectively. 
    Then, we can obtain an $\epsilon$-approximation of the history state with complexity
    \begin{equation}
        \Or\left(  \frac{\alpha_L+\alpha_H}{\eta } \log\left( \frac{ \alpha_L+\alpha_H }{\eta \epsilon } \right) \log\left( \frac{1}{\epsilon} \right)  \right). 
    \end{equation} 
\end{cor}

\subsection{Generalized heat equation}

Consider the following evolutionary partial differential equation (PDE) on a $d$-dimensional unit cube
\begin{align}
    \partial_t u(t,x) &= a \Delta u(t,x) + b \nabla \cdot u(t,x) + c(t,x) u(t,x) + f(t,x), \quad t \in [0,T], \quad x \in [0,1]^d, \quad \label{eqn:heat_PDE}\\
    u(0,x) &= u_0(x), \quad x \in [0,1]^d, \\
    u(t,x) &= 0, \quad x \in \partial ([0,1]^d). 
\end{align}
Here $a > 0$ is a positive parameter called the thermal diffusivity, $b$ is a real parameter called the flow velocity, $c(t,x) \leq 0$ is the potential function, and $f(t,x)$ is the source term. 
We impose homogeneous Dirichlet boundary condition on the cube, so the heat process is dissipative. 

A standard way of solving~\cref{eqn:heat_PDE} numerically is by the method of lines, which first semi-discretizes the PDE into an ODE by only performing spatial discretization then solves the resulting ODE by any time propagator. 
For spatial discretization, along each direction we use $(n_x+1)$ many equi-distant grid points $j/n_x$, $0 \leq j \leq n_x$. 
We discretize the spatial derivatives by central difference, and the semi-discretized heat equation is given as 
\begin{align}
    \frac{d \vec{u}(t) }{dt} &= a L \vec{u}(t) + b D \vec{u}(t) + C(t) \vec{u}(t) + \vec{f}(t), \label{eqn:heat_ODE}\\
    \vec{u}(0) &= \vec{u}_0. 
\end{align}
Here $\vec{u}(t)$ is an $(n_x+1)^d$-dimensional vector approximating the solution $u(t,x)$ at the point $x = (j_1/n_x,j_2/n_x,\cdots,j_d/n_x)$, $0 \leq j_k \leq n_x$. 
$\vec{f}(t)$ is a vector encoding $f(t,x)$ at those points as well. 
$L$, $D$ and $C(t)$ are the discretized version of the Laplace operator $\Delta$, the divergence operator $\nabla\cdot$, and the potential function $c(t,x)$, respectively. 
They are given as 
\begin{equation}
    L = \sum_{j=1}^d I^{\otimes (j-1)} \otimes L_1 \otimes I^{\otimes (d-j)}, \quad L_1 = n_x^2 \left( \begin{array}{cccccc}
        -2 & 1 & & & & \\
         1&-2 &1 & & &\\
         & 1 & -2 & 1 & &\\
         & & \ddots & \ddots &\ddots &\\
         & & &1 & -2 & 1\\
         & & & & 1 & -2
    \end{array} \right), 
\end{equation}
\begin{equation}
    D = \sum_{j=1}^d I^{\otimes (j-1)} \otimes D_1 \otimes I^{\otimes (d-j)}, \quad D_1 = \frac{1}{2} n_x \left( \begin{array}{cccccc}
         0 & 1 & & & & \\
         -1& 0 &1 & & &\\
         & -1 & 0 & 1 & &\\
         & & \ddots & \ddots &\ddots &\\
         & & & -1 & 0 & 1\\
         & & & & -1 & 0
    \end{array} \right), 
\end{equation}
and 
\begin{equation}
    C(t) = \text{diag}( c(t,x_j) ), \quad x_j = (j_1/n_x,j_2/n_x,\cdots,j_d/n_x), \quad 0 \leq j_k \leq n_x. 
\end{equation}
We assume that $n_x$ is sufficiently large such that the spatial discretization error is already reasonably bounded, and we only consider applying quantum algorithm to approximate the semi-discretized solution $\vec{u}(t)$ in~\cref{eqn:heat_ODE}. 

The coefficient matrix of~\cref{eqn:heat_ODE} is $A(t) = aL + bD + C(t)$ and has the following properties. 
First, both $L$ and $C(t)$ are Hermitian, and $D$ is anti-Hermitian. 
The matrix $C(t)$ is negative semi-definite since $c(t,x) \leq 0$,  and eigenvalues of $L_1$ are given as $\lambda = - 4 n_x^2 \left( \sin\left( \frac{j \pi}{ 2(n_x+2) } \right) \right)^2  $, $1 \leq j \leq n_x + 1$. 
So, we have 
\begin{equation}
    A(t) + A(t)^{\dagger} = 2aL + 2C(t) \leq - 8 a d n_x^2 \left( \sin\left( \frac{\pi}{ 2(n_x+2) } \right) \right)^2 \leq -  \frac{8 a d n_x^2}{ (n_x+2)^2 }, 
\end{equation}
where we have used the fact that  $\sin(\frac{\pi x}{2}) \geq x$ for $x \in [0, 1],$
and thus~\cref{eqn:heat_ODE} satisfies our assumption for fast-forwarding by choosing $\eta \sim d$. 
We also have $\norm{A(t)} = \Or( d n_x^2 )$, supposing $c(t,x)$ is on $\Or(1)$. 
So, we have the following complexity estimate for solving~\cref{eqn:heat_ODE} from~\cref{cor:Dyson_hist_inhomo}. 

\begin{cor}\label{cor:heat}
    Consider using the truncated Dyson series method for solving the semi-discretized heat equation in~\cref{eqn:heat_ODE}. 
    Let $\epsilon > 0$ be the target error, $T$ be the evolution time, $d$ be the spatial dimension and $n_x$ be the number of spatial grid points along each dimension. 
    Then, we can obtain an $\epsilon$-approximation of 
    \begin{enumerate}
        \item the history state with complexity
    \begin{equation}
        \widetilde{\Or} \left( n_x^2  \log\left( T \right)  \left( \log\left(\frac{1}{\epsilon}\right)\right)^2 \right), 
    \end{equation}
        \item the final state with complexity 
    \begin{equation}
        \widetilde{\Or}\left( \frac{\max_t \norm{\vec{u}(t)}}{\norm{\vec{u}(T)}} n_x^2 \sqrt{ d T}  \left( \log\left( \frac{ 1 } {\epsilon} \right)\right)^2 \right). 
    \end{equation}
    \end{enumerate}
\end{cor}

When $b = 0$, $c \equiv 0$ and $f \equiv 0$,~\cref{eqn:heat_PDE} represents the standard heat equation. A detailed study on the quantum and classical algorithms for heat equations has been carried out in~\cite{LindenMontanaroShao2022}. 
Notice that the problem setups of~\cite{LindenMontanaroShao2022} and our work are different: our work aims to prepare a quantum state encoding the solution, while~\cite{LindenMontanaroShao2022} studies an estimation of the total heat over a subdomain. 
It would be interesting to further investigate the complexity of measuring the total heat using our fast-forwarded algorithm as a subroutine, and our work is likely to yield a better time dependence compared to~\cite{LindenMontanaroShao2022} for the linear-equations based algorithms. 

\section*{Acknowledgements}

We thank Andrew Childs, David Jennings, and Matteo Lostaglio for helpful discussions.


\bibliographystyle{quantum}
\bibliography{ref}

\clearpage

\appendix

\section{Proof of \texorpdfstring{\cref{lem:stability_continuous_ODE}}{}}\label{app:proof_stability}

\begin{proof}[Proof of~\cref{lem:stability_continuous_ODE}]
    Let $\Delta t = t_1-t_0$. 
    We first prove the upper bound. 
    We use the convergence of the first-order time-dependent Trotter formula and obtain 
    \begin{align}
        \mathcal{T} e^{\int_{t_0}^{t_1} A(s) ds} = \lim_{r\to \infty} \prod_{j=0}^{r-1} e^{ \frac{\Delta t}{r} A(t_0+j\Delta t/r) }. 
    \end{align}
    For each fixed integer $r$ and $0 \leq j \leq r-1$, by the definition of $\eta$, we have 
    \begin{equation}
        \norm{ e^{ \frac{\Delta t}{r} A(t_0 + j\Delta t/r) } } \leq e^{-\eta \Delta t /r}, 
    \end{equation}
    so 
    \begin{align}
        \norm{ \mathcal{T} e^{\int_{t_0}^{t_1} A(s) ds} } \leq \lim_{r \to \infty} \prod_{j=0}^{r-1} \norm{e^{ \frac{\Delta t}{r} A(t_0+j\Delta t/r) }} \leq \lim_{r \to \infty} \prod_{j=0}^{r-1} e^{-\eta \Delta t/r}  = e^{-\eta \Delta t}. 
    \end{align}

    For the lower bound, notice that $ \left(\mathcal{T} e^{\int_{t_0}^{t_1} A(s) ds }\right)^{-1} = \mathcal{T} e^{-\int_{t_0}^{t_1} A(t_0+t_1-s) ds}$. 
    We have 
    \begin{equation}
        1 = \norm{\ket{v}} = \norm{  \left(\mathcal{T} e^{-\int_{t_0}^{t_1} A(t_0+t_1-s) ds}\right)\left(\mathcal{T} e^{\int_{t_0}^{t_1} A(s) ds }\right) \ket{v} } \leq \norm{\mathcal{T} e^{-\int_{t_0}^{t_1} A(t_0+t_1-s) ds}} \norm{\mathcal{T} e^{\int_{t_0}^{t_1} A(s) ds } \ket{v} }, 
    \end{equation}
    and thus 
    \begin{equation}
        \norm{\mathcal{T} e^{\int_{t_0}^{t_1} A(s) ds } \ket{v} } \geq \norm{\mathcal{T} e^{-\int_{t_0}^{t_1} A(t_0+t_1-s) ds}}^{-1}. 
    \end{equation}
    It suffices to derive an upper bound of $\mathcal{T} e^{-\int_{t_0}^{t_1} A(t_0+t_1-s) ds}$, which can be achieved by the similar technique as in the first part of the proof. 
    Specifically, by the Trotter formula again, we have 
    \begin{equation}
        \mathcal{T} e^{-\int_{t_0}^{t_1} A(t_0+t_1-s) ds} = \lim_{r \to \infty} \prod_{j=0}^{r-1} e^{ - \frac{\Delta t}{r} A(t_1-j\Delta t/r) }.  
    \end{equation}
    Noticing that $\norm{e^{ - \frac{\Delta t}{r} A(t_1-j\Delta t/r) }} \leq e^{(\Delta t/r) \max_t\norm{A(t)}}$ for every $r$ and $0 \leq j \leq r-1$, we have 
    \begin{equation}
        \norm{\mathcal{T} e^{-\int_{t_0}^{t_1} A(t_0+t_1-s) ds}} \leq \lim_{r \to \infty} \prod_{j=0}^{r-1} \norm{ e^{ - \frac{\Delta t}{r} A(t_1-j\Delta t/r) } } \leq \lim_{r \to \infty} \prod_{j=0}^{r-1} e^{(\Delta t/r) \max_t\norm{A(t)}} = e^{\Delta t \max_t\norm{A(t)}}. 
    \end{equation}
    Therefore 
    \begin{equation}
        \norm{\mathcal{T} e^{\int_{t_0}^{t_1} A(s) ds } \ket{v} } \geq \norm{\mathcal{T} e^{-\int_{t_0}^{t_1} A(t_0+t_1-s) ds}}^{-1} \geq e^{-\Delta t \max_t\norm{A(t)}}. 
    \end{equation}
\end{proof}

\section{Proof of \texorpdfstring{\cref{lem:2_norm_block_bound}}{}}\label{app:proof_2norm_block_bound}

\begin{proof}[Proof of~\cref{lem:2_norm_block_bound}]
    The idea is that, for any matrix $\mathbf{M}$, its spectral norm is the square root of the spectral norm of the Hermitian matrix $\mathbf{M}^{\dagger}\mathbf{M}$, and the spectral norm of a Hermitian matrix is bounded by its ``$1$-norm'' in terms of the blocks. 
    
    Specifically, we write $\mathbf{M}^{\dagger}\mathbf{M} = \left( B_{i,j} \right)$, where 
    \begin{equation}
        B_{i,j} = \sum_{k=0}^{n-1} M^{\dagger}_{k,i} M_{k,j}. 
    \end{equation}
    Let $\lambda$ be the largest eigenvalue of $\mathbf{M}^{\dagger} \mathbf{M}$, and $\mathbf{x} = [x_0;x_1;\cdots;x_{n-1}]$ be its eigenvector. 
    Suppose that $q = \mathrm{argmax}_j \|x_j\|$, and by the definition of an eigenvector we have 
    \begin{equation}
        \sum_{j=0}^{n-1} B_{q,j} x_j = \lambda x_q. 
    \end{equation}
    By triangle inequality, 
    \begin{equation}
        \lambda \norm{x_q} \leq \sum_{j=0}^{n-1} \norm{B_{q,j}} \norm{x_j}, 
    \end{equation}
    and 
    \begin{equation}
        \lambda \leq \sum_{j=0}^{n-1} \norm{B_{q,j}} \frac{\norm{x_j}}{\norm{x_q}} \leq \sum_{j=0}^{n-1} \norm{B_{q,j}} \leq \max_i \sum_{j=0}^{n-1} \norm{B_{i,j}}. 
    \end{equation}
    Therefore 
    \begin{equation}
        \|\mathbf{M}\| = \sqrt{\lambda} \leq \max_i \sqrt{\sum_{j=0}^{n-1} \norm{B_{i,j}}} \leq \max_i \sqrt{\sum_{j=0}^{n-1} \sum_{k=0}^{n-1} \norm{M_{k,i}} \norm{M_{k,j}} } = \max_i \sqrt{\sum_{k=0}^{n-1} \sum_{j=0}^{n-1} \norm{M_{k,i}} \norm{M_{k,j}} }. 
    \end{equation}
    For each fixed $i$ and $k$, we have 
    \begin{equation}
        \sum_{j=0}^{n-1} \norm{M_{k,i}} \norm{M_{k,j}} = \norm{M_{k,i}} \left( \sum_{j=0}^{n-1} \norm{M_{k,j}} \right) \leq \norm{M_{k,i}} \left( \max_l \sum_{j=0}^{n-1} \norm{M_{l,j}} \right). 
    \end{equation}
    So 
    \begin{equation}
        \norm{\mathbf{M}} \leq \max_i \sqrt{ \sum_{k=0}^{n-1} \left( \norm{M_{k,i}} \left( \max_l \sum_{j=0}^{n-1} \norm{M_{l,j}} \right) \right) } = \sqrt{ \left( \max_i  \sum_{k=0}^{n-1}  \norm{M_{k,i}}\right) \left( \max_l  \sum_{j=0}^{n-1} \norm{M_{l,j}} \right)  }, 
    \end{equation}
    which is the claimed inequality despite different notation of indices. 
\end{proof}

\section{Proof of \texorpdfstring{\cref{thm:complexity_single_step_homo}}{}}\label{app:proof_complexity_single_step_homo}

\begin{proof}[Proof of~\cref{thm:complexity_single_step_homo}]
    The idea is the same as proving~\cref{thm:complexity_single_step_inhomo} but with tighter upper bound for the numerical discretization error. 
    For the ease of reading, here we still present a complete proof, which has some overlap with the proof of~\cref{thm:complexity_single_step_inhomo} but can be read independently. 
    
    Let $M = T/h$, $\mathbf{u} = \mathbf{A}_{M,0}^{-1} \mathbf{b}_{M,0}$ and $\ket{\mathbf{u}} = \mathbf{u} / \norm{\mathbf{u}}$, and we denote $L_{-1} = I$. 
    We first show that $\ket{\mathbf{u}}$ is indeed an $\epsilon/2$-approximation of $\ket{\mathbf{u}_{\text{exact}}} $. 
    Let 
    \begin{equation}\label{eqn:proof_complexity_def_epsilon_p}
        \epsilon' = \frac{ \eta^{3/2} h \epsilon }{ 32 \sqrt{  \max \norm{A(t)}} }.
    \end{equation}
    Now the local truncation error is bounded as 
    \begin{equation}
        \norm{L_j^{-1} R_j - \mathcal{T} e^{\int_{jh}^{(j+1)h} A(s) ds}  } \leq \epsilon'. 
    \end{equation}
    Then, for each $1 \leq k \leq T/h$, we have 
    \begin{align}
        \norm{u_k - u(kh)} &= \norm{ \prod_{j=0}^{k-1} L_j^{-1} R_j u_0 - \mathcal{T} e^{\int_{0}^{kh} A(s) ds}  u_0 } \\
        & \leq \norm{ \prod_{j=0}^{k-1} L_j^{-1} R_j  - \mathcal{T} e^{\int_{0}^{kh} A(s) ds}  } \norm{u_0} \\
        & = \norm{ \prod_{j=0}^{k-1} P_j  -   \prod_{j=0}^{k-1} Q_j } \norm{u_0} , 
    \end{align}
    where we denote $P_j = L_j^{-1} R_j$ and $Q_j = \mathcal{T} e^{\int_{jh}^{(j+1)h} A(s) ds}$. 
    We use the triangle inequality to write 
    \begin{align}
        \norm{u_k - u(kh)} & \leq \norm{ \sum_{l=0}^{k-1} \left[ \left(\prod_{j=l+1}^{k-1}P_j\right) (P_l - Q_l) \left(\prod_{j=0}^{l-1} Q_j\right) \right] } \norm{u_0} \\
        & \leq \sum_{l=0}^{k-1} \left[ \left(\prod_{j=l+1}^{k-1}\norm{P_j}\right) \norm{P_l - Q_l} \left(\prod_{j=0}^{l-1} \norm{Q_j} \right) \right] \norm{u_0} \\
        & \leq \epsilon' \sum_{l=0}^{k-1} \left(\prod_{j=l+1}^{k-1}\norm{P_j}\right) \left(\prod_{j=0}^{l-1} \norm{Q_j} \right) \norm{u_0} . 
    \end{align}
    By~\cref{lem:stability_continuous_ODE} and the proof of~\cref{lemma:condition-number}, we have $\norm{Q_j} \leq e^{-\eta h} \leq e^{-\eta h/2}$ and $\norm{P_j} \leq e^{-\eta h/2}$, and thus 
    \begin{equation}
        \norm{u_k - u(kh)} \leq k \epsilon' e^{-\eta (k-1)h/2} \norm{u_0} . 
    \end{equation}
    Therefore, the error in the unnormalized history vector can be bounded as 
    \begin{align}
        & \quad \norm{\mathbf{u} - \mathbf{u}_{\text{exact}}} \\
        &= \sqrt{ \sum_{k=1}^{M} \norm{ u_k - u(kh) }^2  } \\
        & \leq \norm{u_0} \epsilon' \sqrt{ \sum_{k=1}^{M} k^2 e^{-\eta (k-1)h} } \\
        & = \norm{u_0} \epsilon' \sqrt{ \frac{(2M^2 + 2M -1)e^{-\eta(M+1)h} - M^2 e^{-\eta(M+2)h} - (M+1)^2e^{-\eta Mh} + e^{-\eta h} + 1}{(1-e^{-\eta h})^3} } \\
        & = \norm{u_0} \epsilon' \sqrt{ \frac{1+e^{-\eta h}}{(1-e^{-\eta h})^3} - \frac{M^2 e^{-\eta Mh} ( 1- e^{-\eta h } )^2 }{(1-e^{-\eta h})^3} -  \frac{ 2Me^{-\eta Mh} (1-e^{-\eta h}) }{(1-e^{-\eta h})^3} - \frac{e^{-\eta M h}(1+e^{-\eta h}) }{(1-e^{-\eta h})^3} } \\
        & \leq \norm{u_0} \epsilon' \sqrt{ \frac{2}{(1-e^{-\eta h})^3} } \\
        & \leq 4 \frac{\norm{u_0} \epsilon'}{ (\eta h)^{3/2} }.   
    \end{align}
    The error in the quantum states becomes 
    \begin{equation}\label{eqn:proof_complexity_single_homo_eq1}
         \norm{\ket{\mathbf{u}} - \ket{\mathbf{u}_{\text{exact}}} } \leq \frac{2}{\norm{\mathbf{u}_{\text{exact}}} } \norm{\mathbf{u} - \mathbf{u}_{\text{exact}}} \leq 8 \frac{\norm{u_0} \epsilon'}{ \norm{\mathbf{u}_{\text{exact}}} (\eta h)^{3/2} }. 
    \end{equation}
    According to the lower bound in~\cref{lem:stability_continuous_ODE}, the norm of the exact unnormalized history vector satisfies 
    \begin{align}
        \norm{\mathbf{u}_{\text{exact}}} &= \sqrt{ \sum_{j=0}^{M} \norm{ u(jh) }^2 } \\
        & = \sqrt{ \sum_{j=0}^{M} \norm{ \mathcal{T} e^{\int_{0}^{jh} A(s) ds} u_0 }^2 } \\
        & \geq \norm{u_0} \sqrt{ \sum_{j=0}^{M} e^{- 2 j h \max \norm{A(t)}  } }  \\
        & = \norm{u_0} \sqrt{ \frac{1 - e^{- 2 (T+h) \max \norm{A(t)} } }{1 - e^{-2h \max \norm{A(t)}}} }  \\
        & \geq \frac{\norm{u_0} }{2 \sqrt{ h \max \norm{A(t)}} }. 
    \end{align}
    Then~\cref{eqn:proof_complexity_single_homo_eq1} becomes 
    \begin{equation}
         \norm{\mathbf{u} - \mathbf{u}_{\text{exact}}} \leq 16 \frac{\sqrt{  \max \norm{A(t)}} \epsilon'}{ h \eta^{3/2} } \leq \frac{\epsilon}{2}, 
    \end{equation}
    where the last inequality is due to the choice of $\epsilon'$ in~\cref{eqn:proof_complexity_def_epsilon_p}. 

    Then, solving the linear system of equations $\mathbf{A}_{M,0} \mathbf{u} = \mathbf{b}_{M,0}$ up to error $\epsilon/2$ gives the $\epsilon$-approximation of $\ket{\mathbf{u}_{\text{exact}}}$. 
    According to~\cite{CostaAnYuvalEtAl2022}, it requires $\Or( \kappa \log(1/\epsilon) )$ queries to the block-encoding of $\mathbf{A}_{M,0}$ and the state preparation oracle of $\mathbf{b}_{M,0}$, where $\kappa = \norm{\mathbf{A}_{M,0}} \norm{\mathbf{A}_{M,0}^{-1}}$ is the condition number of $\mathbf{A}_{M,0}$. 
    According to~\cref{lemma:condition-number}, we have 
    \begin{equation}
        \kappa = \Or\left( \left( 1 + \max_j \norm{L_j} + \max_j \norm{R_j} \right) \left(1+\max_j\norm{L_j^{-1}} \right)  \frac{M}{\eta T}  \right)
    \end{equation}
    and complete the proof. 
\end{proof}

\section{Proof of \texorpdfstring{\cref{cor:Dyson_hist_homo}}{}}\label{app:proof_Dyson_hist_homo}

\begin{proof}[Proof of~\cref{cor:Dyson_hist_homo}]
    According to~\cite[Section 3]{BerryCosta2022}, constructing a block-encoding of $\mathbf{A}_{M,0}$ needs $K$ calls to $O_A$. 
    Also notice that $L_j = I$ in the truncated Dyson series method. 
    Then~\cref{thm:complexity_single_step_homo} tells that the overall query complexity is 
    \begin{equation}\label{eqn:proof_Dyson_eq4}
        \Or\left( \left( 1 + \max_j \norm{R_j} \right)  \frac{KM}{\eta T} \log\left( \frac{1}{\epsilon} \right)  \right)
    \end{equation}
    queries to $O_A$ and 
    \begin{equation}\label{eqn:proof_Dyson_eq5}
        \Or\left( \left( 1  + \max_j \norm{R_j} \right)  \frac{M}{\eta T} \log\left( \frac{1}{\epsilon} \right)  \right)
    \end{equation}
    queries to $O_{u_0}$. 

    We need to choose $K$ and $M$ such that~\cref{eqn:num_assump_complexity_homo} holds. 
    Notice that, under the assumption $\eta h \leq 1$ which will be satisfied by the choice of $h$ specified later, we have 
    \begin{equation}
        \eta h e^{-\eta h} = \Omega\left( \eta h \right) = \Omega\left( \frac{\eta^{3/2} h}{ \sqrt{\max\norm{A(t)}} } \right) = \Omega\left( \frac{\eta^{3/2} h \epsilon }{ \sqrt{\max\norm{A(t)}} } \right). 
    \end{equation}
    Then it suffices to choose $K$ and $M$ such that 
    \begin{equation}
        \norm{ R_j - \mathcal{T} e^{\int_{jh}^{(j+1)h} A(s) ds}  } \leq \Or\left( \epsilon' \right)
    \end{equation}
    with 
    \begin{equation}\label{eqn:proof_Dyson_epsilon_p}
        \epsilon' = \frac{\eta^{3/2} h \epsilon }{ \sqrt{\max\norm{A(t)}} }. 
    \end{equation}

    We first choose 
    \begin{equation}\label{eqn:proof_Dyson_eq1}
        M = 2 \alpha_A T, 
    \end{equation}
    then $\alpha_A h = 1/2$ and thus the block-encoding of the Dyson series is well-behaved~\cite{BerryCosta2022}. 
    As a result, by the definition of $R_j$ in~\cref{eqn:Dyson_R}, we have 
    \begin{equation}\label{eqn:proof_Dyson_eq2}
        \norm{R_j} \leq \sum_{k=0}^K \frac{h^k}{k!} \alpha_A^k \leq e^{\alpha_A h} = \Or(1). 
    \end{equation}
    Furthermore, according to~\cite[Eq.(6)]{BerryCosta2022}, the local truncation error can be bounded as 
    \begin{equation}
         \norm{ R_j - \mathcal{T} e^{\int_{jh}^{(j+1)h} A(s) ds}  } \leq \Or\left( \frac{ (\alpha_A h)^{K+1} }{(K+1)!} \right) \leq \Or\left( \frac{ 1 }{(K+1)!} \right) \leq  \Or\left( e^{-K} \right). 
    \end{equation}
    Therefore, to bound the local truncation error by $\epsilon'$, it suffices to choose $K = \Or\left( \log(1/\epsilon') \right)$, which in turn, according to the choice of $\epsilon'$ in~\cref{eqn:proof_Dyson_epsilon_p}, becomes 
    \begin{equation}\label{eqn:proof_Dyson_eq3}
        K = \Or\left(  \log\left( \frac{\max\norm{A(t)}}{\eta h \epsilon } \right)  \right) = \Or\left(  \log\left( \frac{ \alpha_A }{\eta \epsilon } \right)  \right). 
    \end{equation}
    Plugging~\cref{eqn:proof_Dyson_eq1,eqn:proof_Dyson_eq2,eqn:proof_Dyson_eq3} back to~\cref{eqn:proof_Dyson_eq4,eqn:proof_Dyson_eq5} yields the claimed complexity estimates. 
\end{proof}

\section{Constructing block-encodings in lower-order methods}
\label{appendix:block-encodings}

In order to design a quantum algorithm, it is crucial to determine the input model.
Recall~\cref{eqn:def_oracle_A},
here we mainly follow the paradigm in~\cite{BerryCosta2022}, by assuming we can query $A(t)$ in the following way:
\begin{equation}
    O_A \ket{0^a} \ket{t} \ket{\psi} = \frac{1}{\alpha_A} \ket{0^a} \ket{t} A(t) \ket{\psi} + \ket{\perp}.     
\end{equation}
Here $a$ is the number of ancilla qubits, and the first register is used to indicate a successful or failed measurement. $\ket{t}$ ranges from $\ket{0}$ to $\ket{M-1}$, representing a uniform grid on the time interval $[0,T]$. $\ket{\perp}$ represents a possibly unnormalized state such that $\left(\ket{0^a}\bra{0^a} \otimes I\right) \ket{\perp} = 0 $, and $\alpha_A \geq \max \norm{A(t)}$ is the block-encoding factor. 

Because of the definition of ~\cref{eqn:def_A_padding}, we could encounter the scenario that the clock register may contain
some state that is beyond time $T$.
However we could pay the price of at most $M_p$ controlled-$X$ gate acting on the
first register controlled on the clock register is greater than $\ket{M-1}$, so the definition in~\cref{eqn:def_oracle_A} will
not be influenced.

Another useful operator is the $\mathrm{ADD}$ operator, namely
\begin{equation}
    O_{\mathrm{ADD}} := \sum_{t=0}^{M+M_p-1}\ket{(t+1) \text{ mod } (M+M_p-1)}\bra{t}.
\end{equation}

\subsection{Forward Euler method}
\begin{lem}
    When using the Forward Euler method,
    we are able to construct a block-encoding of the matrix $\mathbf{A}_{M,M_{p}-1}$ defined in~\cref{eqn:def_A_padding} with block-encoding factor $2+h\alpha_A$, 
    using $1$ query to $O_A$, $2$ queries to the $\mathrm{ADD}$ operator, 
    $2$ queries to the preparation oracles defined in~\cref{eqn:forward-euler-prep-orcale}, and at most
    $M_p + 1$ multi-qubit controlled-X gate.
\end{lem}
\begin{proof}
    
The Forward Euler method is defined as follows:
\begin{equation}
    u_{j+1} = \left(I + h A(jh)\right) u_j.
\end{equation}
So in this scenario we have
\begin{equation}
    L_j = I, \quad R_j = I + hA(jh).
\end{equation}
Thus the matrix $\mathbf{A}_{M,M_p-1}$ becomes
\begin{equation}
\begin{split}
    \mathbf{A}_{M,M_p-1} &= 
    \begin{bmatrix}
        I & & & & & & & & \\
        -I - hA(0) & I & & & \\
         & -I - hA(h) & I & & \\
         & & \ddots &\ddots & & & & &  \\
         & & & -I - hA((M-1)h) & I\\
         & & & & -I & I & & & \\
         & & & &  & -I & I & & \\ 
         & & & &  &  & \ddots & \ddots & \\ 
         & & & &  & & & -I & I \\ 
     \end{bmatrix}\\
    &=
    \mathbb{I} + 
    \begin{bmatrix}
        0 & & & & & & & & \\
        -I - hA(0) & 0 & & & \\
         & -I - hA(h) & 0 & & \\
         & & \ddots &\ddots & & & & &  \\
         & & & -I - hA((M-1)h) & 0\\
         & & & & -I & 0 & & & \\
         & & & &  & -I & 0 & & \\ 
         & & & &  &  & \ddots & \ddots & \\ 
         & & & &  & & & -I & 0 \\ 
     \end{bmatrix}\\
    &=
    \mathbb{I} - \sum_{t=0}^{M+M_p-2}\ket{t+1}\bra{t} \otimes I - 
    \begin{bmatrix}
        0 & & & & & & & & \\
        - hA(0) & 0 & & & \\
         & - hA(h) & 0 & & \\
         & & \ddots &\ddots & & & & &  \\
         & & & - hA((M-1)h) & 0\\
         & & & & 0 & 0 & & & \\
         & & & &  & 0 & 0 & & \\ 
         & & & &  &  & \ddots & \ddots & \\ 
         & & & &  & & & 0 & 0 \\ 
     \end{bmatrix}\\
    &=
    \mathbb{I} - \sum_{t=0}^{M+M_p-2}\ket{t+1}\bra{t} \otimes I - h\sum_{t=0}^{M-1} \ket{t+1}\bra{t} \otimes A(t).
\end{split}
\end{equation}
Let us now consider the queries needed for constructing a block-encoding of the $\mathbf{A}_{M+M_p-1}$ above. 
Notice that
\begin{equation}
\label{eqn:block-encoding-sub-diag-I}
   (I \otimes O_{\mathrm{ADD}} \otimes I)(\ket{0^a}\ket{t}\ket{\psi})
   = \ket{0^a}\ket{(t+1) \text{ mod } (M+M_p-1)}\ket{\psi},
\end{equation}
then controlled on the second register being $\ket{0}$, flip the first register to $\ket{1}\ket{0^{a-1}}$,
we then have a block-encoding of $\sum_{j=0}^{M-1}\ket{j+1}\bra{j} \otimes I$.
Denote this block-encoding as $U_2$.

Similarly, 
\begin{equation}
   (I \otimes O_{\mathrm{ADD}} \otimes I)O_A(\ket{0^a}\ket{t}\ket{\psi})
   = \frac{1}{\alpha_A}\ket{0^a}\ket{t+1}A(t)\ket{\psi} + \ket{\perp},
\end{equation}
thus we have $U_3$, which is a block-encoding of $h\sum_{j=0}^{M-1} \ket{j+1}\bra{j} \otimes A(jh)$ with block-encoding factor $\alpha_A h$. 
By leveraging the linear combination of unitaries technique~\cite{ChildsWiebe2012}, 
we need two additional ancilla qubits. 
We construct the preparation oracle as
\begin{equation}\label{eqn:forward-euler-prep-orcale}
    O_{\mathrm{prep}} = 
    \frac{1}{\sqrt{2 + h\alpha_A}}
    \begin{bmatrix}
        1 & * & * & *\\
        \imath & * &* & *\\
        \imath \sqrt{\alpha_A h} &* &* & *\\
        0 &* &* & *\\
    \end{bmatrix}
\end{equation}
and the select oracle to be
\begin{equation}
    O_{\mathrm{sel}} = \ket{01}\bra{01} \otimes U_2 + \ket{10}\bra{10} \otimes U_3.
\end{equation}
Simple computation yields
\begin{equation}
    (\overline{O_{\mathrm{prep}}}\otimes I_{a} \otimes \mathbb{I})^\dag O_{\mathrm{sel}} (O_{\mathrm{prep}}\otimes I_{a} \otimes \mathbb{I}) = 
    \ket{00}\bra{00}\frac{1}{2 + h\alpha_A}  \left(I_{a}\otimes \mathbb{I} - U_2 - (\alpha_A h)U_3\right) + P^{\perp}.
\end{equation}
Here $\overline{O_{\mathrm{prep}}}$ is the conjugate of $O_{\mathrm{prep}}$, and
$P^{\perp}$ satisfies $(\bra{00} \otimes I_a \otimes \mathbb{I})P^{\perp}(\ket{00} \otimes I_a \otimes \mathbb{I}) = 0$.
Thus we are able to have a block-encoding for $\mathbf{A}_{M,M_p-1}$ with block-encoding factor $2+h\alpha_A$. 
\end{proof}

\subsection{Trapezoidal rule}
\begin{lem}
    When using the trapezoidal rule,
    we are able to construct a block-encoding of the matrix $\mathbf{A}_{M,M_{p}-1}$ defined in~\cref{eqn:def_A_padding} with block-encoding factor $2+h\alpha_A$, 
    using $2$ queries to $O_A$, $2$ queries to the $\mathrm{ADD}$ operator,
    $2$ queries to the preparation oracles defined in~\cref{eqn:forward-euler-prep-orcale}, and at most
    $2M_p + 1$ multi-qubit controlled-X gate.
\end{lem}
\begin{proof}
For the trapezoidal rule, we have
\begin{equation}
    u_{j+1} = u_j + \frac{1}{2}h(A(jh)u_j + b(jh) + A((j+1)h)u_{j+1} + b((j+1)h)).
\end{equation}
\begin{equation}
    \left(I - \frac{h}{2}A((j+1)h)\right)u_{j+1} = \left(I + \frac{h}{2}A(jh)\right)u_j + \frac{h}{2}(b(jh)+b((j+1)h)).
\end{equation}
So we would have
\begin{equation}
    L_j = I - \frac{h}{2}A((j+1)h),\quad
    R_j = I + \frac{h}{2}A(jh).
\end{equation}

Thus the matrix $\mathbf{A}_{M,M_p-1}$ becomes
\begin{equation}
\begin{split}
    &\mathbf{A}_{M,M_p-1} \\
    &= 
    \begin{bmatrix}
        I & & & & \\
        -I - \frac{h}{2}A(0) & I - \frac{h}{2}A(h) & & & \\
         & -I - \frac{h}{2}A(h) & I - \frac{h}{2}A(2h) & & \\
         & & \ddots &\ddots & \\
         & & & -I - \frac{h}{2}A((M-1)h) & I - \frac{h}{2}A(Mh)\\
         & & & & -I & I & & & \\
         & & & &  & -I & I & & \\ 
         & & & &  &  & \ddots & \ddots & \\ 
         & & & &  & & & -I & I \\ 
     \end{bmatrix}\\
    &=
    \mathbb{I} + 
    \begin{bmatrix}
        0 & & & & \\
        -I - \frac{h}{2}A(0) & - \frac{h}{2}A(h) & & & \\
         & -I - \frac{h}{2}A(h) & - \frac{h}{2}A(2h) & & \\
         & & \ddots &\ddots & \\
         & & & -I - \frac{h}{2}A((M-1)h) & - \frac{h}{2}A(Mh)\\
         & & & & -I & 0 & & & \\
         & & & &  & -I & 0 & & \\ 
         & & & &  &  & \ddots & \ddots & \\ 
         & & & &  & & & -I & 0 \\ 
     \end{bmatrix}\\
    &=
    \mathbb{I} - \sum_{t=0}^{M+M_p-2}\ket{t+1}\bra{t} \otimes I - 
    \begin{bmatrix}
        0 & & & & \\
         - \frac{h}{2}A(0) & - \frac{h}{2}A(h) & & & \\
         &  - \frac{h}{2}A(h) & - \frac{h}{2}A(2h) & & \\
         & & \ddots &\ddots & \\
         & & &  - \frac{h}{2}A((M-1)h) & - \frac{h}{2}A(Mh)\\
         & & & & 0 & 0 & & & \\
         & & & &  & 0 & 0 & & \\ 
         & & & &  &  & \ddots & \ddots & \\ 
         & & & &  & & & 0 & 0 \\ 
     \end{bmatrix}\\
    &=
    \mathbb{I} - \sum_{t=0}^{M+M_p-2}\ket{t+1}\bra{t} \otimes I 
    - \frac{h}{2}\sum_{t=0}^{M-1} \ket{t+1}\bra{t} \otimes A(t) - \frac{h}{2}\sum_{t=1}^{M} \ket{t}\bra{t} \otimes A(t).
\end{split}
\end{equation}

The rest is pretty similar to the process in the forward Euler part.
Recall~\cref{eqn:block-encoding-sub-diag-I},
we have a block-encoding $V_2$ of $\sum_{t=0}^{M+M_p-2}\ket{t+1}\bra{t}\otimes I$ using one query to $O_{\mathrm{ADD}}$ and one multi-qubit controlled-X gate.

Similarly, we have a block-encoding $V_3$ of $\frac{h}{2}\sum_{t=0}^{M-1} \ket{t+1}\bra{t} \otimes A(t)$ with block-encoding factor $\frac{1}{2}h\alpha_A$
and a block-encoding $V_4$ of $\frac{h}{2}\sum_{t=1}^{M} \ket{t}\bra{t} \otimes A(t)$ with block-encoding factor $\frac{1}{2}h\alpha_A$. 
Constructing them will cost us at most $2M_p$ multi-qubit controlled-X gate additionally.

By leveraging the linear combination of unitaries technique~\cite{ChildsWiebe2012}, 
we will need two additional ancilla qubits.
We construct the preparation oracle as
\begin{equation}\label{eqn:trapezoidal-prep-orcale}
    O_{\mathrm{prep}} = 
    \frac{1}{\sqrt{2 + h\alpha_A}}
    \begin{bmatrix}
        1 & * & * & *\\
        \imath & * &* & *\\
        \imath \sqrt{\alpha_A h/2} &* &* & *\\
        \imath \sqrt{\alpha_A h/2} &* &* & *\\
    \end{bmatrix}
\end{equation}
and the select oracle to be
\begin{equation}
    O_{\mathrm{sel}} = \ket{01}\bra{01} \otimes V_2 + \ket{10}\bra{10} \otimes V_3 + \ket{11}\bra{11} \otimes V_4.
\end{equation}
Simple computation yields
\begin{equation}
    (\overline{O_{\mathrm{prep}}}\otimes I_{a} \otimes \mathbb{I})^\dag O_{\mathrm{sel}} (O_{\mathrm{prep}}\otimes I_{a} \otimes \mathbb{I}) = 
    \ket{00}\bra{00}\frac{1}{2 + h\alpha_A}  \left(I_{a}\otimes \mathbb{I} - V_2 -\frac{\alpha_A h}{2} V_3 -\frac{\alpha_A h}{2} V_4\right) + P^{\perp}.
\end{equation}
Here $\overline{O_{\mathrm{prep}}}$ is the conjugate of $O_{\mathrm{prep}}$, and
$P^{\perp}$ satisfies $(\bra{00} \otimes I_a \otimes \mathbb{I})P^{\perp}(\ket{00} \otimes I_a \otimes \mathbb{I}) = 0$.
Thus we are able to have a block-encoding for $\mathbf{A}_{M,M_p-1}$ with block-encoding factor $2+h\alpha_A$.
\end{proof}

\section{Error analysis of lower-order numerical schemes}
\label{appendix:ea}
In the following, we demonstrate that Forward Euler and Trapezoidal rule
have local error scale as $\mathcal{O}(h^2)$ and $\mathcal{O}(h^3)$ respectively, when approximating the time-ordering matrix exponential. For simplicity, we only consider the time interval $[0,h]$, where $h$ is the step size. Furthermore, 
we would require that $A(t)$ is in $C^3$.
\subsection{Forward Euler}
\subsubsection{Evolution operator}
Since
\begin{equation}
\begin{split}
    \left\|\mathcal{T}e^{\int_0^h A(s)\mathrm{d}s} - (I + hA(0))\right\| &= 
    \sup_{\|v\|= 1}\left\|\left(\mathcal{T}e^{\int_0^h A(s)\mathrm{d}s} - (I + hA(0))\right)v\right\|. 
\end{split}
\end{equation}

Let us now consider 
\begin{equation}
\begin{split}
    \frac{\mathrm{d}u}{\mathrm{d}t} &= A(t) u,\\
    u(0) &= v,
\end{split}
\end{equation}
we can then denote $\mathcal{T}e^{\int_0^h A(s)\mathrm{d}s} v$ as $u(h)$.
\begin{equation}
\label{eqn:fe-estimate}
\begin{split}
    \sup_{\|v\| = 1} \left\|u(h) - (I + hA(0))v\right\|
    &= \sup_{\|v\| = 1}\left\|v + A(0)h + \int_0^h \frac{\mathrm{d}^2u}{\mathrm{d}t^2} t\; \mathrm{d}t - (I + hA(0))v\right\|\\
    &= \sup_{\|v\| = 1} \left\|\int_0^h \frac{\mathrm{d}^2u}{\mathrm{d}t^2} t\; \mathrm{d}t\right\| \leq 
    \sup_{\|v\| = 1} \sup_{0 \leq t \leq h} \frac{h^2}{2}\norm{\frac{\mathrm{d}^2 u}{\mathrm{d}t^2}}.
\end{split}
\end{equation}
Notice that 
\begin{equation}
\begin{split}
    \frac{\mathrm{d}^2u}{\mathrm{d}t^2} &= \frac{\mathrm{d}}{\mathrm{d}t} (A(t) u(t)) = A'(t) u(t) + A^2(t)u(t),
\end{split}
\end{equation}
so we have the estimation
\begin{equation}
\label{eqn:fe-estimate2}
    \sup_{\|v\| = 1}\sup_{0\leq t \leq h}\left\|\frac{\mathrm{d}^2 u}{\mathrm{d}t^2}\right\| 
    = \sup_{\|v\| = 1}\sup_{0\leq t \leq h}\|A'(t) u(t) + A^2(t)u(t)\| \leq \sup_{0\leq t \leq h}(\|A(t)\|^2 + \|A'(t)\|).
\end{equation}
Plugging~\cref{eqn:fe-estimate2} back into~\cref{eqn:fe-estimate}, we have
\begin{equation}
    \left\|\mathcal{T}e^{\int_0^h A(s)\mathrm{d}s} - (I + hA(0))\right\| \leq \frac{1}{2}\sup_{0\leq t \leq h}(\|A(t)\|^2 + \|A'(t)\|) h^2 = \Or(h^2). 
\end{equation}

\subsubsection{Inhomogeneous term}
Here, we want  to bound
\begin{equation}
\begin{split}
    \left\|h b(h) - \int_0^h\mathcal{T}e^{\int_s^h A(t)\mathrm{d}t}b(s)\mathrm{d}s\right\| = \mathcal{O}(h^2).
\end{split}
\end{equation}
Since 
\begin{equation}
    f(s)|_{s = h} := \mathcal{T}e^{\int_s^h A(t)\mathrm{d}t}b(s)a\bigg{|}_{s = h} = b(h),
\end{equation}
then we have
\begin{equation}
    \int_0^h f(s) \mathrm{d} s - h f(h) = \int_0^h f(h) + (s-h)f'(h) + \mathcal{O}((s-h)^2)\;\mathrm{d}s - hf(h) = \mathcal{O}(h^2).
\end{equation}

\subsection{Trapezoidal rule}
\label{appendix:ea-tr}
\subsubsection{Evolution operator}
For the trapezoidal rule, we are estimating
\begin{equation}
\label{eqn:tr-estimate}
\begin{split}
    \left\|\mathcal{T}e^{\int_0^h A(s)\mathrm{d}s} - (I - \frac{h}{2}A(h))^{-1}(I + \frac{h}{2}A(0))\right\| = 
    \sup_{\|v\|= 1}\left\|\left(\mathcal{T}e^{\int_0^h A(s)\mathrm{d}s} - (I - \frac{h}{2}A(h))^{-1}(I + \frac{h}{2}A(0))\right)v\right\|. 
\end{split}
\end{equation}
According to the Neumann series, we know that
\begin{equation}
\label{eqn:tr-update}
\begin{split}
    &(I - \frac{h}{2}A(h))^{-1}(I + \frac{h}{2}A(0)) = 
    \sum_{k=0}^\infty (\frac{h}{2}A(h))^k (I + \frac{h}{2}A(0))\\
    &= I + \frac{h}{2} A(0) + \frac{h}{2}A(h) + \frac{h^2}{4}A(h)A(0) +
    \frac{h^2}{4}A^2(h) + \mathcal{O}(h^3).
\end{split}
\end{equation}
Now, since 
\begin{equation}
\begin{split}
    A(h) = A(0) + hA'(0) + \frac{h^2}{2}A''(0) + \mathcal{O}(h^3),
\end{split}
\end{equation}
we can further write~\cref{eqn:tr-update} as
\begin{equation}
\label{eqn:tr-update-analysis}
  (I - \frac{h}{2}A(h))^{-1}(I + \frac{h}{2}A(0)) =   I + hA(0) + \frac{h^2}{2}A'(0) + \frac{h^2}{2}A^2(0) + \mathcal{O}(h^3).
\end{equation}

Next, using a similar notation in the above section, we know that
\begin{equation}
\begin{split}
    &\sup_{\|v\| = 1} \left\|u(h) - (I - \frac{h}{2}A(h))^{-1}(I + \frac{h}{2}A(0))v\right\|\\
    &= \sup_{\|v\| = 1}\left\|v + hu'(0) + \frac{h^2}{2}u''(0) + \frac{1}{2}\int_0^h \frac{\mathrm{d}^3u}{\mathrm{d}t^3} t^2\; \mathrm{d}t - (I - \frac{h}{2}A(h))^{-1}(I + \frac{h}{2}A(0))v\right\|.
\end{split}
\end{equation}
Notice that
\begin{equation}
    u'(t) = A(t) u(t),
\end{equation}
\begin{equation}
    u''(t) = \frac{\mathrm{d}}{\mathrm{d}t} u'(t) = \frac{\mathrm{d}}{\mathrm{d}t} A(t) u(t)= A'(t) u(t) + A^2(t) u(t),
\end{equation}
\begin{equation}
    u'''(t) = \frac{\mathrm{d}}{\mathrm{d}t} u''(t) = \frac{\mathrm{d}}{\mathrm{d}t} (A'(t) u(t) + A^2(t) u(t))= A''(t) u(t) + A'(t) A(t) u(t) + 
    2A(t)A'(t) u(t) + A^3(t) u(t),
\end{equation}
we know
\begin{equation}
\label{eqn:exact-update-tr}
    v + hu'(0) + \frac{h^2}{2}u''(0) = (I + hA(0) + \frac{h^2}{2}(A'(0) + A^2(0)))v,
\end{equation}
Compare~\cref{eqn:exact-update-tr} and~\cref{eqn:tr-update-analysis}
and plug everything in then equation~\cref{eqn:tr-estimate} could be $\mathcal{O}(h^3)$.
\subsubsection{Inhomogeneous term}
Here we need to bound
\begin{equation}
\begin{split}
    \left\|(I - \frac{h}{2}A(h))^{-1}\frac{h}{2}(b(h) + b(0)) - \int_0^h\mathcal{T}e^{\int_s^h A(t)\mathrm{d}t}b(s)\mathrm{d}s\right\| = \mathcal{O}(h^3), 
\end{split}
\end{equation}
which can be achieved by the following computation
\begin{equation}
\begin{split}
    &\quad \left\|\left(I - \frac{h}{2}A(h))^{-1}\frac{h}{2}(b(h) + b(0))\right) - \int_0^h\mathcal{T}e^{\int_s^h A(t)\mathrm{d}t}b(s)\mathrm{d}s\right\|\\
    &=
    \left\|\left(I + \frac{h}{2}A(h) + \mathcal{O}(h^2)\right)\frac{h}{2}(b(h) + b(0)) - \int_0^h\mathcal{T}e^{\int_s^h A(t)\mathrm{d}t}b(s)\mathrm{d}s\right\|\\
    &\leq
    \left\|\left(I + \frac{h}{2}A(h) + \mathcal{O}(h^2)\right)\frac{h}{2}(b(h) + b(0)) - \frac{h}{2}\left(b(h) + e^{\int_0^h A(t)\mathrm{d}t}b(0)\right)\right\| \\
    & \quad\quad +
    \left\|\frac{h}{2}\left(b(h) + e^{\int_0^h A(t)\mathrm{d}t}b(0)\right) - \int_0^h\mathcal{T}e^{\int_s^h A(t)\mathrm{d}t}b(s)\mathrm{d}s\right\|\\
    &\leq 
    \left\|\left(I + \frac{h}{2}A(h) + \mathcal{O}(h^2)\right)\frac{h}{2}(b(h) + b(0)) - \frac{h}{2}\left(b(h) + e^{\int_0^h A(t)\mathrm{d}t}b(0)\right)\right\|
    +
    \mathcal{O}(h^3)\\
    &=
    \left\|\frac{h^2}{4}A(h) b(h) + \left(I + \frac{h}{2}A(h)\right)\frac{h}{2}b(0) - \frac{h}{2} e^{\int_0^h A(t)\mathrm{d}t}b(0)\right\|
    +
    \mathcal{O}(h^3)\\
    &=
    \left\|\frac{h^2}{4}A(h) b(h) + \left(I + \frac{h}{2}A(h)\right)\frac{h}{2}b(0) - \frac{h}{2} (b(0) + hA(0) b(0))\right\|
    +
    \mathcal{O}(h^3)\\
    &=
    \left\|\frac{h^2}{4}A(0) b(0) + \left(I + \frac{h}{2}A(0)\right)\frac{h}{2}b(0) - \frac{h}{2} (b(0) + hA(0) b(0))\right\|
    +
    \mathcal{O}(h^3) = \mathcal{O}(h^3).
\end{split}
\end{equation}

\end{document}